%% file: metal_cmb_nat.tex
\documentclass[useAMS,usenatbib]{mnras}

\include{include_tex/pacchetti}
\usepackage{url}
\usepackage{color}
\include{include_tex/journals}

\include{include_tex/definizioni}

\include{include_tex/additional_def}

\defcitealias{Pallottini:2014_sim}{P14}
\defcitealias{valliniSUB}{V15}

\begin{document}

\date{}
\pagerange{\pageref{firstpage}--\pageref{lastpage}} \pubyear{2015}
\title[Mapping high-$z$ metals with FIR lines]{Mapping metals at high redshift with far-infrared lines}
\author[Pallottini et al.]{A. Pallottini$^{1}$\thanks{email: andrea.pallottini@sns.it}, S. Gallerani$^{1}$, A. Ferrara$^{1,2}$, B. Yue$^{1}$, L. Vallini$^{3}$, R. Maiolino$^{4}$, \newauthor~\& C. Feruglio$^{1,5}$\\
$^{1}$Scuola Normale Superiore, Piazza dei Cavalieri 7, I-56126 Pisa, Italy\\
$^{2}$Kavli IPMU, The University of Tokyo, 5-1-5 Kashiwanoha, Kashiwa 277-8583, Japan\\
$^{3}$Dipartimento di Fisica e Astronomia, Universit\'{a} di Bologna, viale Berti Pichat 6/2, 40127 Bologna, Italy\\
$^{4}$Cavendish Laboratory, University of Cambridge, 19 J. J. Thomson Ave., Cambridge CB3 0HE, UK\\
$^{5}$INAF, Osservatorio Astronomico di Roma, Via Frascati 33, 00040 Monteporzio Catone, Italy
}

\maketitle

\label{firstpage}

\begin{abstract}
Cosmic metal enrichment is one of the key physical processes regulating galaxy formation and the evolution of the intergalactic medium (IGM). However, determining the metal content of the most distant galaxies has proven so far almost impossible; also, absorption line experiments at $z\gsim 6$ become increasingly difficult because of instrumental limitations and the paucity of background quasars. With the advent of ALMA, far-infrared emission lines provide a novel tool to study early metal enrichment. Among these, the \CII~line at 157.74~$\mu$m is the most luminous line emitted by the interstellar medium of galaxies. It can also resonant scatter CMB photons inducing characteristic intensity fluctuations ($\Delta I/I_{\rm CMB}$) near the peak of the CMB spectrum, thus allowing to probe the low-density IGM.
We compute both \CII~galaxy emission and metal-induced CMB fluctuations at $z\sim 6$ by using Adaptive Mesh Refinement cosmological hydrodynamical simulations and produce mock observations to be directly compared with ALMA BAND6 data ($\nu_{\rm obs}\sim 272$~GHz).
The \CII~line flux is correlated with $M_{\rm UV}$ as $\log(F_{\rm peak}/\mu{\rm Jy})= -27.205 -2.253\,M_{\rm UV} -0.038\,M_{\rm UV}^2$. Such relation is in very good agreement with recent ALMA observations \citep[e.g.][]{Maiolino:2015arXiv,capak:2015arXiv} of $M_{\rm UV}<-20$ galaxies. We predict that a $M_{\rm UV}=-19$ ($M_{\rm UV}=-18$) galaxy can be detected at $4\sigma$ in $\simeq 40$ ($2000$) hours, respectively.
CMB resonant scattering can produce $\simeq\pm 0.1~\mu$Jy/beam emission/absorptions features that are very challenging to be detected with current facilities. The best strategy to detect these signals consists in the stacking of deep ALMA observations pointing fields with known $M_{\rm UV}\simeq -19$ galaxies. This would allow to simultaneously detect both \CII~emission from galactic reionization sources and CMB fluctuations produced by $z\sim 6$ metals.
\end{abstract}


\section{Introduction}\label{sec_intro}

In $\Lambda$CDM cosmologies\footnote{In this work we assume a $\Lambda$CDM cosmology with total matter, vacuum and baryonic densities in units of the critical density $\Omega_{\Lambda}= 0.727$, $\Omega_{dm}= 0.228$, $\Omega_{b}= 0.045$, Hubble constant $\rm H_0=100~h~km~s^{-1}~Mpc^{-1}$ with $\rm h=0.704$, spectral index $n=0.967$, $\sigma_{8}=0.811$ \citep[][]{Larson:2011}.}, structure formation is a hierarchical bottom-up process \citep[i.e.][]{Press:1974}. Dwarf galaxies are expected to be abundant and to represent the first ($z\gsim10$) efficient metal factories of the Universe \citep[e.g.][]{Madau:2001ApJ,Ferrara:2008IAUS}. These galaxies effectively pollute their surrounding intergalactic medium (IGM), since -- compared to larger galaxies -- dwarfs have a shallower potential well and because their smaller size allows multiple SN to coherently drive the outflows \citep[e.g.][]{Ferrara:2000MNRAS}.

In spite of the impressive progresses produced by deep optical/IR surveys \citep{Dunlop13,Madau14,Bouwens:2014arXiv} in identifying galaxies well within the Epoch of Reionization, very little is known about the metallicity and other properties of these systems, including feedback \citep[e.g.][]{Dayal14} and interactions with their environment \citep[e.g.][]{Barnes:2014PASP}.

The situation is slightly better for what concerns the IGM, where metal enrichment is typically studied by measuring the abundance of heavy elements at different cosmic times with quasar (QSO) absorption line spectroscopy \citep[e.g.][]{Songaila:2005AJ,Ryan-Weber:2009MNRAS,Becker:2009ApJ,Simcoe:2011ApJ,DOdorico:2013MNRAS} and -- more recently -- with gamma-ray bursts soft X-ray absorption \citep[e.g.][]{campana:2010MNRAS,Behar:2011ApJ,Campana:2015Aa}. These observations show that the diffuse IGM is enriched at metallicity $Z\gsim 10^{-3.5}\zsun$ at any overdensity ($\Delta$) and redshift ($z$) probed so far \citep{Meiksin:2009RvMP}. As the gas left over by the Big Bang is (virtually) metal-free, understanding how and when the first metals were produced holds the key of many structure formation processes.

As a consequence of the paucity of standard luminous lighthouses prior to reionization \citep{Barkana:2001PhR,Ciardi:2005SSR}, the enrichment at $z\gsim6$ cannot be efficiently studied with absorption line experiments. Additionally, at high-$z$ the detection of most metal tracers becomes increasingly difficult. For example, \CIV~metal absorption line detections at high-$z$ are rare since the \CIV~doublet shifts into the near-infrared. In this spectral regions observations become challenging because of increased OH emission from the sky and severe telluric absorption.

The above situation is going to change very soon thanks to the advent of the Atacama Large Millimeter-submillimeter Array (ALMA). This observatory gives us access to far-infrared (FIR) and sub-mm spectroscopy, where metal lines carrying information about the energetics of the interstellar medium and outflows are found. FIR emission lines are a powerful tool to study high-$z$ galaxies, since -- unlike Ly$\alpha$ -- they are affected neither by the increasing IGM neutral fraction at $z\gsim6$ nor by the presence of dust.

In particular, the \CII~{\small$\left(^{2}P_{3/2} \rightarrow\,^{2}P_{1/2}\right)$} line at 157.74~$\mu$m is the brightest FIR emission line, i.e. it can account for $\sim1\%$ of the total infrared luminosity of galaxies \citep[e.g.][]{Crawford:1985ApJ,Madden:1997ApJ}. Before the advent of ALMA, \CII~observations at $z\gsim4$ were limited to QSO host galaxies and to rare galaxies with extreme star formation ($\simeq10^3\msun\,{\rm yr}^{-1}$)\citep[e.g.][]{maiolino:2005AA,Carilli:2013ARA&A,gallerani:2012aa,cicone:2015aa}. The unprecedented resolution and sensitivity of ALMA offer the unique opportunity to search for \CII~emission from normal star forming galaxies ($\simeq1-10\msun\,{\rm yr}^{-1}$) at high-$z$ \citep[e.g.][]{Maiolino:2015arXiv,capak:2015arXiv}. These observations yield key information about the stellar age, gas metallicity, star formation, and also on the environments in which these early galaxies formed \citep[e.g.][]{GonzalezLopez:2014ApJ,DeLooze:2014AA}. Recently, the \CII~emission arising from high-$z$ star forming galaxies has been modelled through numerical simulations \citep[e.g.][]{nagamine:2006ApJ,Vallini:2013MNRAS,tomassetti:2015MNRAS,valliniSUB} and analytical models \citep[e.g.][]{Gong:2012ApJ,munoz:2014MNRAS}.

In addition to tracing the ISM of galaxies, FIR resonant lines correspond to transitions that are excited via a resonant scattering process of Comic Microwave Background (CMB) photons on heavy element atoms present in the IGM \citep[e.g.][]{deBernardis:1993,Maoli:1996ApJ}. These resonant transitions are available for essentially all the most abundant species, e.g. C, N, O, Si, S and Fe, in the mid-IR and FIR wavelength ranges \citep[e.g.][]{Basu:2004na}.

As a result, IGM metals can produce CMB spectral distortions and spatial fluctuations that can be used to extract unique information on the cosmic metal enrichment process, like, for example, spatial distribution maps of a given metal species. This experiment is conceptually similar to QSO absorption line studies, but it uses the CMB as a background source: the method has the enormous advantage that every pixel in a sky map can act as a source against which metal lines may appear either in emission or absorption.

Various works \citep[e.g.][]{Maoli:2005ESASP,Basu:2007NewAR,Schleicher:2008AA} have proposed CMB fluctuations/distortions as a tool to trace the smooth distribution of metals and molecules in the post-recombination Universe ($z\gg10$). These fluctuations typically affect the first CMB multipoles -- at $l\lsim10^{2}$, large angular scales ($\theta\sim 1 \,{\rm deg}$) -- and they will be hopefully seen by \emph{Planck} successors \citep[e.g.][]{hernandeza:2006ApJ,Chluba:2014arXiv}. On smaller scales -- $\theta\sim 10\arcsec$ ($l\gsim10^{5}$) -- CMB fluctuations in the FIR can be used to map the enriched IGM surrounding the first galaxies ($z\gsim10$).

Here we present a model based on detailed Adaptive Mesh Refinement hydrodynamical cosmological simulations, that allows us to simultaneously compute both the \CII~line emission from early galaxies and the expected level and statistical properties of CMB fluctuations arising from intergalactic \CIIion~ resonant scattering.

The paper is organized as follows. In Sec. \ref{sec_model}, we describe our model, introducing the adopted high-$z$ cosmic metal enrichment simulation (Sec. \ref{sec_simulazione}). We postprocess the simulation to account for \CII~galaxy emission (Sec. \ref{sec_gal_emission}) and CMB fluctuations (Sec. \ref{sec_cmb_model}), and analyze the properties of the simulated CMB fluctuations in Sec. \ref{sec_cmb_th_anal}. Then, we construct and analyze mock ALMA observations of \CII~galaxy emission and CMB fluctuations (Sec. \ref{sez_mock}). Finally, we propose and discuss an observational strategy for the detection of \CII~galaxy emission and metal-induced CMB fluctuations (Sec. \ref{sec_strategy}). Conclusions are given in Sec. \ref{sec_conclusions}.

\section{Model}\label{sec_model}
\subsection{Cosmological simulations}\label{sec_simulazione}

\begin{figure*}
\centering
\includegraphics[width=0.49\textwidth]{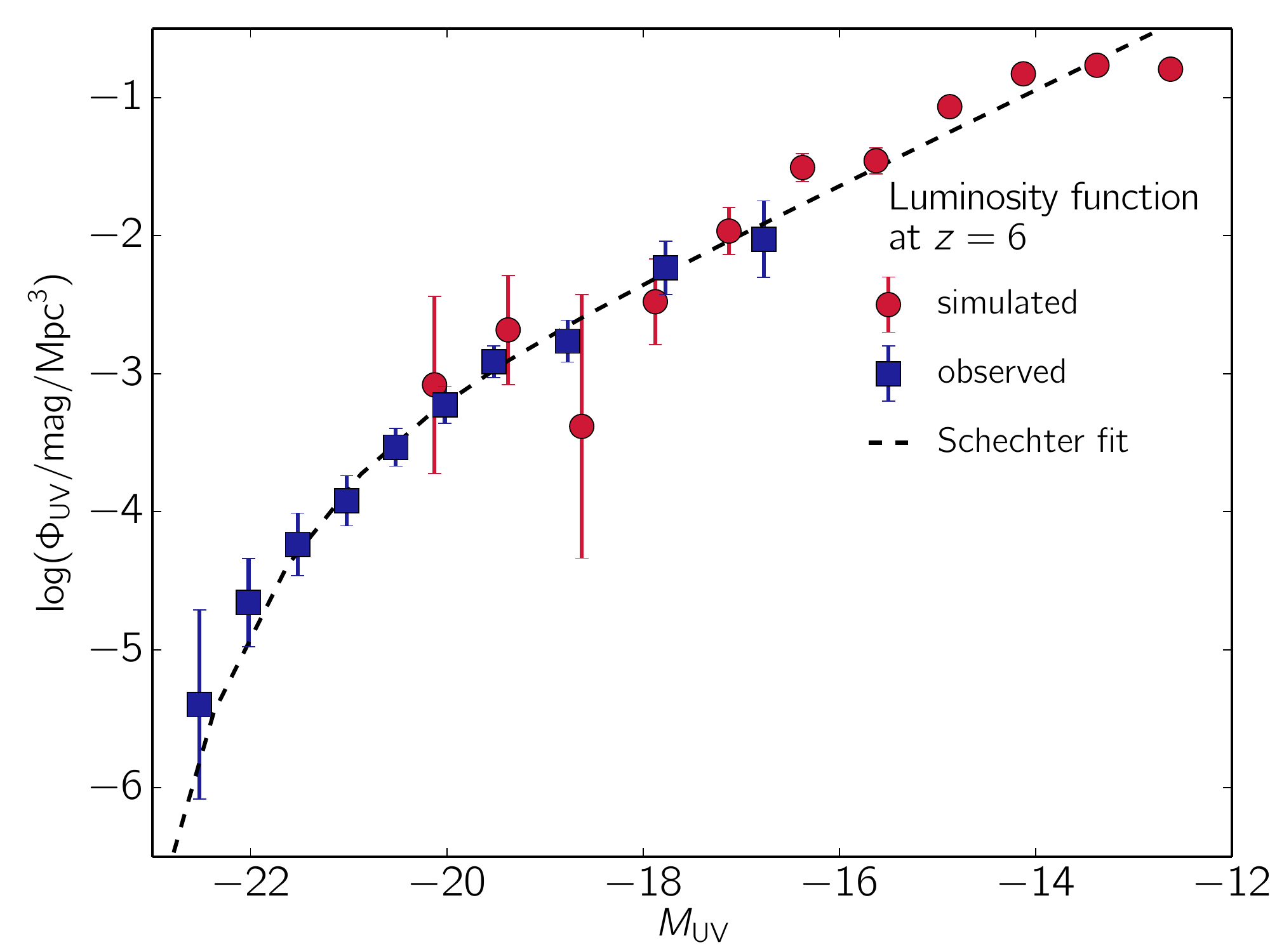}
\includegraphics[width=0.49\textwidth]{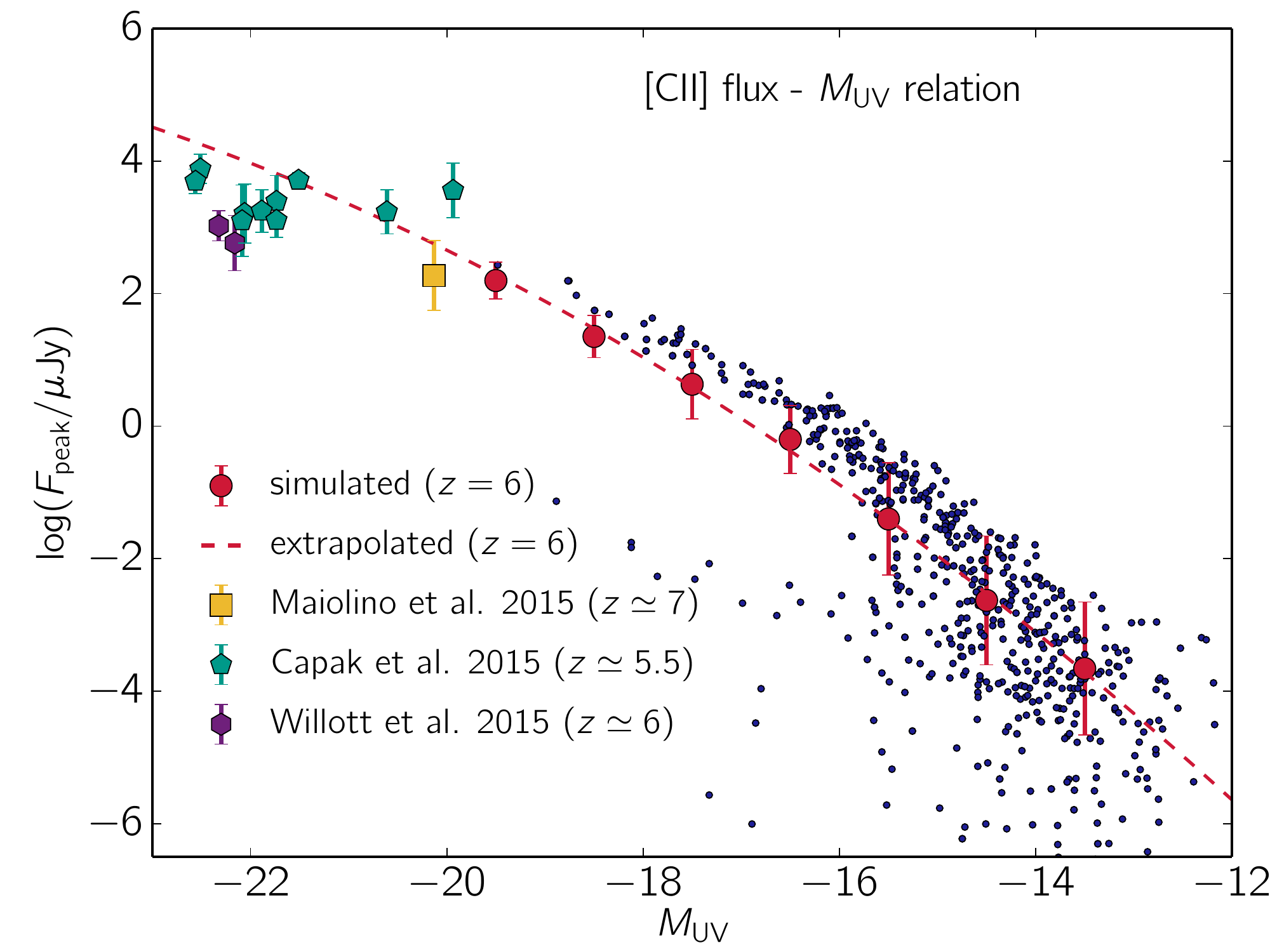}
\caption{
{\bf Left panel}: UV luminosity function, $\Phi_{\rm UV}$ at $z=6$. Red circles represent the simulated $\Phi_{\rm UV}$; blue squares denote the observed $\Phi_{\rm UV}$, inferred from a composite collection of HST datasets \citep{Bouwens:2014arXiv}; black dashed lines are the Schechter fit \citep{Bouwens:2014arXiv} to the observations. {\bf Right panel}: \CII~flux as a function of $M_{\rm UV}$ at $z=6$ (blue small circles); the same relation binned in $\Delta M_{\rm UV}=1$ intervals is shown by red circles. The red errorbars correspond to the r.m.s. variance within the magnitude intervals; The red dashed line is the best fit to the relation. Also plotted are the data from recent high-$z$ observations from \citet{Maiolino:2015arXiv,capak:2015arXiv,Willott:2015arXiv15} with yellow squares, green pentagons and violet hexagons, respectively.
\label{fig_luminosity}
}
\end{figure*}

We start from the suite of cosmological hydrodynamical metal enrichment simulations presented by \citet[][\citetalias{Pallottini:2014_sim} hereafter]{Pallottini:2014_sim}. Using a customized version of the Adaptive Mesh Refinement code \textlcsc{ramses} \citep[][]{Teyssier:2002}, \citetalias{Pallottini:2014_sim} follows the evolution of a $(10$~Mpc~$h^{-1})^{3}$ comoving cosmic volume from $z=199$ until $z=4$ with $512^{3}$ dark matter particles. The gas evolution is tracked on a corresponding number of coarse grid cells, with 4 additional levels of refinement based on a Lagrangian mass threshold-based criterion. This set-up allows us to reach a maximum resolution of $1.22\,h^{-1}$~kpc for the gas in the densest regions.

Star formation and supernova feedback are included via customized subgrid prescriptions; \citetalias{Pallottini:2014_sim} accounts for stellar yields based on population synthesized models \citep[][]{Salvadori:2008MNRAS}. The simulated galaxy sample reproduces the observed cosmic star formation rate \citep[][]{Bouwens:2012ApJ,Zheng:2012Natur} and stellar mass densities \citep[][]{Gonzalez:2011} evolution in the full redshift range $4\leq z \lsim 10$.

In \citetalias{Pallottini:2014_sim}, we track the progressive metal enrichment of the gas, initially of primordial (BBN) composition, and its thermal history by accounting for heating and cooling processes \citep[][]{Theuns:1998MNRAS}. The gas ionization state is regulated by an external, redshift-dependent ionizing UV background \citep[][UVB]{Haardt:1996,Haardt:2012} produced both by stars and QSOs. At $z\simeq6$, the observationally allowed rage for the UV photoionization rate is $\simeq ( 0.9 - 3.6 )\times 10^{-13} {\rm s}^{-1}$ \citep[e.g.][]{Wyithe:2011MNRAS}. We have verified that the gas thermodynamical evolution is only marginally affected by UVB intensity variations within this range.

To compute the ionization state of the various atomic species, we postprocess the simulation outputs using the photoionization code \textlcsc{cloudy} \citep[][]{cloudy:1998PASP}. Similar to \citetalias{Pallottini:2014_sim}, we account both for the UVB intensity at 912 A ($J_\nu$), and the density ($n$), kinetic temperature ($T_{k}$), and metallicity ($Z$) of the gas.

We adopt the following classification for the baryonic gas: (i) the interstellar medium (ISM) defines highly overdense gas ($n\slash n_{\rm mean}=\Delta\geq 10^3$); (ii) the {\it true} intergalactic medium (IGM) is characterized by $\Delta \leq 10$; (iii) the circumgalactic medium (CGM) represents the interface between the IGM and ISM ($10<\Delta<10^3$). \citet[][]{Pallottini:2014cgmh} have shown that the adopted classification is consistent with the one proposed by \citet[][]{2014ApJ...784..142S}, based on the distance $r$ between the gas and the center of the galaxy. Specifically (i) the ISM is located within the galaxy virial radius ($r_{\rm vir}$), (ii) the IGM at $r/r_{\rm vir}\gsim 5$ and (iii) the CGM at $1\lsim r/r_{\rm vir}\lsim 5$.

In \citetalias{Pallottini:2014_sim}, we showed that galaxies develop a stellar mass-metallicity ($M_{\star}$-$Z_{\star}$) relation by $z=6$. For $M_{\star}>10^{7}\msun$, the stellar metallicity ($Z_{\star}$) increases with increasing stellar mass, while for $M_{\star}\lsim10^{7}\msun$, $Z_{\star}$ is constant. This means that, while massive galaxies are able to retain most of their metals, low mass galaxies are prone to metal ejection because of their swallower potential well. This indicates that at high-$z$ low mass galaxies are the main driver of metal enrichment in the IGM \citep[e.g.][]{Ferrara:2008IAUS}. In agreement with previous numerical studies \citep[e.g.][]{Dayal:2013MNRAS}, the $M_{\star}$-$Z_{\star}$ relation shows little evolution from $z=6$ to $z=4$, for galaxies with $M_{\star}\simeq10^{7}$. Our results are consistent with the metallicity evolution suggested by recent observations of $3<z<5$ star forming galaxies \citep[][]{Troncoso:2013arXiv1311}.

We find that (i) most of the gas resides in the IGM, (ii) the denser CGM contains about $15\%$ of the baryons and (iii) the ISM accounts only for a small fraction of the total mass ($\simeq 7\%$). On the other hand, at any given redshift the enriched gas is mostly found near the metal production site, i.e. in the ISM; less than $10\%$ in mass of the produced metals can reach the IGM/CGM. In particular, at $z=4$, a $\Delta$-$Z$ relation is in place: the IGM shows an uniform distribution around $Z\simeq 10^{-3.5}\zsun$, in the CGM $Z$ steeply rises with $\Delta$ up to $10^{-2}\zsun$ and the ISM has $Z\simeq 10^{-1}\zsun$. A considerable fraction ($\gsim50\%$) of the enriched IGM/CGM is in an hot state ($T_{k}\gsim 10^{4.5}{\rm K}$), meaning that most of the carbon is in \CIV. Our analysis shows that \CIV~absorption line experiments can only probe $\simeq2\%$ of the total produced carbon.

\citetalias{Pallottini:2014_sim} results are in agreement with previous numerical studies in terms of the baryon thermodynamical state \citep[e.g.][]{Rasera:2006,Cen:2011ApJ}, the evolution of the metal filling factor \citep[e.g.][]{Oppenheimer:2009MNRAS,Johnson:2013MNRAS} and the $\Delta$-$Z$ relation \citep[e.g.][]{Gnedin:1997,Oppenheimer:2012MNRAS}. Additionally, the preliminary analysis of synthetic spectra extracted from the simulation is consistent with recent observations of metal absorption lines \citep[][]{DOdorico:2013MNRAS}.

As a further consistency check, we test the UV luminosity function ($\Phi_{\rm UV}$) extracted from the \citetalias{Pallottini:2014_sim} simulation against observations of $z=6$ galaxies. For each galaxy in our simulation, $M_{\rm UV}$ is calculated with \textlcsc{starburst99} \citep[][]{starburst99:2010ApJS}, using $Z_{\star}$ and the stellar age as input parameters. In the left panel of Fig. \ref{fig_luminosity}, we plot the simulated and observed $\Phi_{\rm UV}$ with red circles and blue squares, respectively. The observations are taken from a composite collection of HST datasets \citep{Bouwens:2014arXiv}, and we additionally plot the Schechter fit to the observations with a black dashed line. The simulated $\Phi_{\rm UV}$ well matches the observed UV luminosity function for $-20\lsim M_{\rm UV}\lsim-16$. Additionally, the simulated $\Phi_{\rm UV}$ is in fair agreement with the Schechter fit extrapolated at $-16\lsim M_{\rm UV}\lsim-12$. Note that, for the simulated $\Phi_{\rm UV}$, the bright end is around $M_{\rm UV}\sim-19$, where the large scatter is due to the limited simulated volume.

\subsection{\CII~emission from the ISM of high-$z$ galaxies}\label{sec_gal_emission}

Starting from the \citetalias{Pallottini:2014_sim} simulations, we compute the \CII~ emission at 157.74~$\mu {\rm m}$ due to the {\small$^{2}P_{3/2} \rightarrow\,^{2}P_{1/2} $} forbidden transition of ionized carbon in the ISM of galaxies. The \CII~line is the dominant coolant of the galaxy interstellar medium, and it can be collisionally excited under conditions present in different ISM phases, e.g. in the cold and warm neutral medium (CNM, WNM), in high density photodissociation regions (PDRs), and even in the ionized gas \citep[][]{Tielens:1985ApJ,Wolfire:1995ApJ,Wolfire:2003ApJ,Abel:2006MNRAS,Vallini:2013MNRAS}.

In our cosmological hydrodynamic simulations, we cannot resolve sub-kpc scales, that are typical for different ISM structures (CNM, WNM and PDRs). To overcome this problem, we account for \CII~emission\footnote{In the present work, we focus on \CII~emission from $z\simeq6$ galaxies. However, we note that the \citetalias{valliniSUB} model can be readily extended to compute the emission of fine structure lines from other heavy elements, such as, e.g. \NII~at 122~$\mu {\rm m}$ and \OI~at 63~$\mu {\rm m}$ \citep[][]{Vallini:2013MNRAS}.} from the ISM of high-$z$ galaxies by adopting the model presented in \citet[][]{Vallini:2013MNRAS} and updated in \citet{valliniSUB} (\citetalias{valliniSUB} hereafter). \citetalias{valliniSUB} predicts the \CII~emission from CNM, WNM and PDRs within individual galaxies as a function of their metallicity ($Z_{\star}$) and star formation rate (SFR). The \citetalias{valliniSUB} model is based on the radiative transfer simulation of a high-$z$ galaxy; it includes a subgrid treatment for computing the thermodynamical equilibrium of the diffuse neutral medium (on $\simeq 60\,\rm{pc}$ scales); it allows to localize molecular clouds within the galaxy (on $\simeq 1-10\,\rm{pc}$ scales) and to compute the corresponding \CII~ emission by means of the PDR code \textlcsc{ucl\_pdr} \citep[][]{ucl_pdr:2005MNRAS}.

\begin{figure}
\centering
\includegraphics[width=0.49\textwidth]{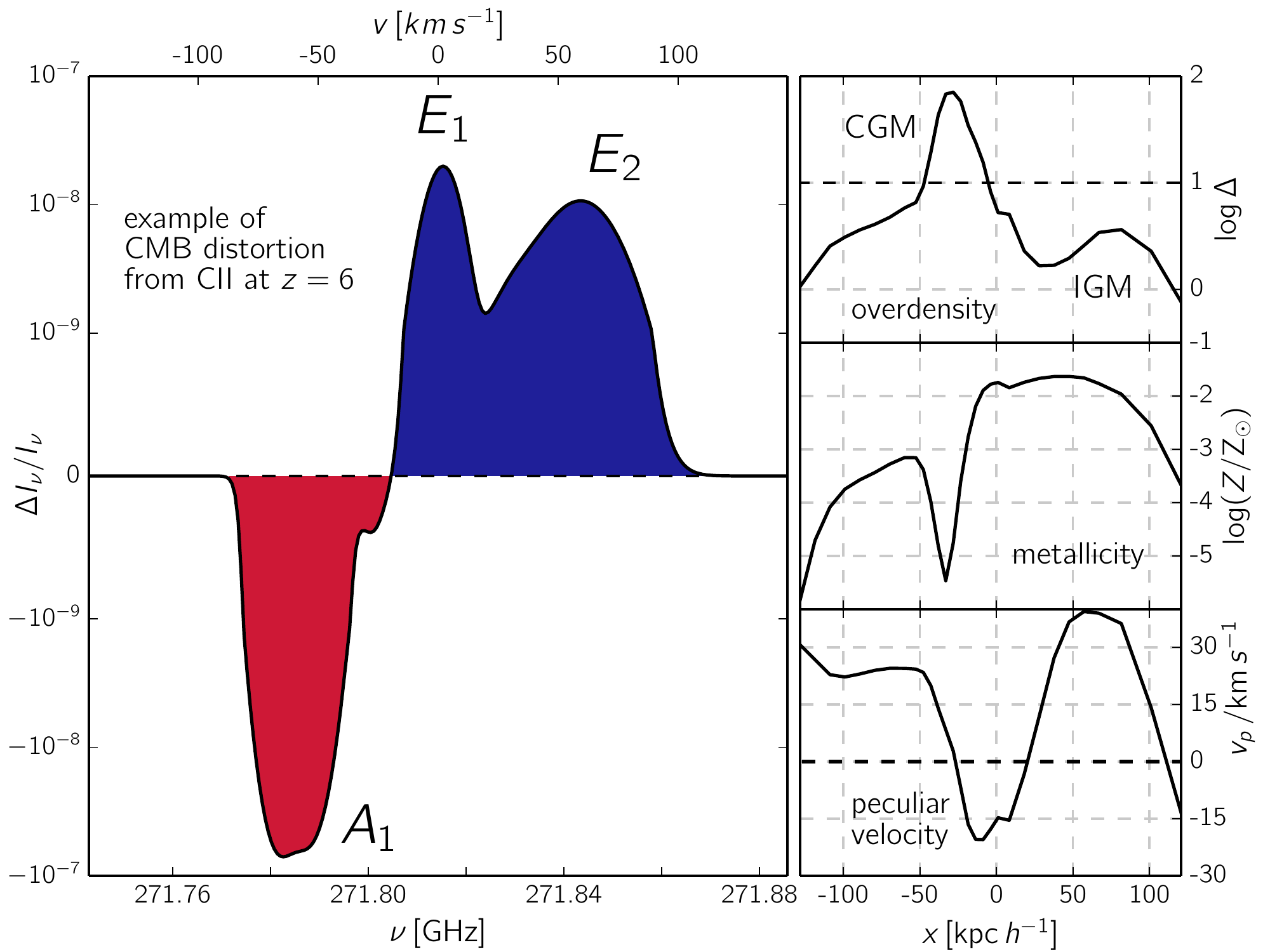}
\caption{\CII~CMB spectral distortions/fluctuations from a simulated line of sight (l.o.s.) at $z=6$. {\bf Left panel}: intensity fluctuations ($\diinu$) as a function of frequency ($\nu$, lower axis) and velocity ($v$, upper axis). Absorption and emission features are highlighted in red and blue, respectively, and the corresponding peaks are labelled with $A_{i}$ and $E_{i}$. Note that integration in the shown bandwidth ($\Delta\nu\simeq0.1$~GHz) yields $\diiDnu\simeq-7\times10^{-7}$. {\bf Right panel}: overdensity ($\Delta$, top), metallicity ($Z$, center) and peculiar velocity ($v_{p}$, bottom) for the \CII~metal patch responsible for the fluctuations ($|\diinu|>0$) as a function of comoving distance (${\rm kpc}\,h^{-1}$). In the overdensity plot our $\Delta$ based IGM/CGM definition is marked with a dashed black line. In the peculiar velocity plot $v_{p}=0$ is marked with a dashed black line. The origins of both the velocity (left panel) and distance (right panel) axis are shifted for displaying purpose.
\label{fig_esempio}}
\end{figure}

According to the \citetalias{valliniSUB} model\footnote{The \citetalias{valliniSUB} model assumes that \HII~regions provide only a negligible contribution to the \CII~emission \citep[][in particular see Fig. 8]{GonzalezLopez:2014ApJ}. However, the relative fraction of \CII~emission arising from \HII regions is not clearly known. As an indicative value, \citet{Vasta:2010MNRAS} observe a median contribution of $\simeq30\%$. For a more detailed discussion, we address the interested reader to \citet[][]{bin:2015mapping}.}, the \CII~galaxy luminosity ($L_{\rm CII}$) depends on the galaxy SFR and metallicity through the following fitting formula:
\begin{align}\label{eq_liv_model}
\log{(L_{\rm CII})}&=7.0+1.2\,\log({\rm SFR}) +0.021\,\log(Z_{\star})\\
	&+0.012\,\log({\rm SFR})\log(Z_{\star})-0.74\,{\log^2(Z_{\star})}\, ,\nonumber
\end{align}
where $L_{\rm CII}$, SFR and $Z_\star$ are in units of $L_\odot$, $\msun\,{\rm yr}^{-1}$ and $\zsun$, respectively. The \citetalias{valliniSUB} model has been used to interpret high-$z$ LAEs and LBGs \CII~observations \citep[e.g.][]{GonzalezLopez:2014ApJ,Ota:2014ApJ}, and \citetalias{valliniSUB} provides predictions consistent with \CII~observations of local metal-poor dwarf galaxies \citep[][]{DeLooze:2014AA}. We compare the \citetalias{valliniSUB} predictions with the work by \citet{munoz:2014MNRAS} (see their eq. 8): for a ${\rm SFR}=10^2 \msun/{\rm yr}$ and a solar metallicity, according to the \citetalias{valliniSUB} model $L_{\rm CII}\simeq2.5\times 10^9 L_\odot$, that is a factor $\simeq5$ higher that the one predicted by \citet{munoz:2014MNRAS}, i.e. $L_{\rm CII}\simeq 5\times 10^{8} L_\odot$. Given the uncertainties on the local \CII-SFR relation, $6\times10^8 \lsim L_{\rm CII}/L_\odot \lsim 10^{10}$, we find that both models are consistent with observations, though the \citetalias{valliniSUB} predictions provide a better match. To compute $L_{\rm CII}$ for galaxies in the \citetalias{Pallottini:2014_sim} simulation we use eq. \ref{eq_liv_model}, where $Z_{\star}$ and ${\rm SFR}$ are the mass-averaged values derived from the simulation \citep[see also][]{bin:2015mapping}.

We calculate the relation between \CII~and UV emission. In the right panel of Fig. \ref{fig_luminosity}, we show the \CII~flux as a function of $M_{\rm UV}$ for the simulated galaxies at $z=6$ and for various $z=5-7$ ALMA observations \citep[][]{Maiolino:2015arXiv,capak:2015arXiv,Willott:2015arXiv15}.
The simulated relation is shown as a scatter plot, binned in $\Delta M_{\rm UV}=1$ intervals. The plot shows a clear correlation between \CII~flux and UV magnitude for the simulated galaxies, with a Spearman correlation coefficient of $\simeq-0.78$. The best fit formula for the relation is
\be\label{eq_flux_fit}
\log(F_{\rm peak}/\mu{\rm Jy})= -27.205 -2.253\,M_{\rm UV} -0.038\,M_{\rm UV}^2.
\ee
Using eq. \ref{eq_flux_fit} we can slightly extrapolate the simulated relation to $M_{\rm UV}<-20$, and directly compare it with recent ALMA observations (right panel of Fig. \ref{fig_luminosity}), i.e. a \CII~detection in a $z\simeq7$ LBG \citep[][yellow square]{Maiolino:2015arXiv}, observations of several \quotes{normal} ($\sim L_{\star}$) galaxies at $z\simeq5-6$ \citep[][green pentagons]{capak:2015arXiv} and 2 LBGs at $z\simeq6$ \citep[][violet hexagons]{Willott:2015arXiv15}. Our model is in good agreement with the observations and local determinations ($z\simeq0$) of the $L_{\rm CII}$-SFR relation \citep[][]{DeLooze:2014AA}, as we discuss in more details in Sec. \ref{sec_strategy}. For simulated galaxies brighter than $M_{\rm UV}<-16$, the $F_{\rm peak} - M_{\rm UV}$ relation shows a bimodal trend: while $90\%$ of the galaxies follow eq. \ref{eq_flux_fit}, the remaining $\simeq 10\%$ have fainter fluxes as a consequence of their lower metallicity values (see eq. \ref{eq_liv_model}).

We finally note that we limit our comparison between simulation results and observational data to $z\simeq6$ normal star forming galaxies (${\rm SFR}\lsim10^2$) without considering \CII~observations in $z\simeq6$ QSO. In the latter case, observations refer to galaxies hosting super massive black holes ($M_\bullet\sim10^{9}\msun$) that are not present in the \citetalias{Pallottini:2014_sim} simulation, because of its limited volume.

\subsection{CMB scattering from intergalactic metals}\label{sec_cmb_model}

The IGM and CGM are not seen in emission, since their typical densities are too low ($n<0.1~{\rm cm}^{-3}$) for the upper levels of the \CII~transition to be efficiently populated through collisions with electrons and/or protons \citep[][]{Suginohara:1999ApJ,Gong:2012ApJ}. In this case the spin temperature of the transition approaches the CMB one\footnote{\citet{Gong:2012ApJ} have shown that the IGM/CGM spin temperature of the \CII~transition is close to the CMB temperature up to $z\simeq2$. At lower $z$, the soft UV background pumping effect may be sufficiently strong to decouple the spin temperature from the CMB temperature. Note that the redshift of the decoupling depend on the considered transition \citep[e.g.][, for \OI]{hernandeza:2007ApJ}.}.

As a result a \CIIion~ion at rest with respect to the CMB cannot show up in emission or absorption against this background \citep[e.g.][]{deBernardis:1993,Maoli:1996ApJ,daCunha:2013ApJ}. However, if the ion has a peculiar velocity, for example approaching with a velocity $v_{\rm p}$, it will receive a larger (i.e. Doppler-boosted) CMB photon flux in the direction of its motion. As CMB photons are resonantly scattered \citep[e.g.][]{Basu:2004na,Basu:2007NewAR,Schleicher:2008AA} by an atomic transition of frequency $\nu_{0}\simeq1901$~GHz the result is an emission feature in the CMB spectrum at $\nu_{\rm obs}={\nu_{0}}/{(1+z)}$. In the case of a receding \CIIion~ion, the opposite situation occurs, resulting into an absorption feature. \CIIion~ions in the IGM and CGM are therefore expected to generate CMB spectral distortions whose intensity depends on both their abundance and peculiar velocity, the latter being of the order of $\sim50\,{\rm km}\,{\rm s}^{-1}$.

The differential amplitude $\Delta I_{\nu}$ of the signal with respect to the CMB intensity $I_{\nu}$ is given by \citep[][]{Maoli:1996ApJ}:
\be\label{eq_maoli}
\diinu=\left(1-e^{-\tau_{\nu}}\right)(3-\alpha_{\nu})(v_{\rm p}/c)\,
\ee
where $\alpha_{\nu}=\nu({\rm d}I_{\nu}/{\rm d}\nu)/I_{\nu}$ is the CMB spectral index and the optical depth ($\tau_{\nu}$) can be written as \citep[e.g.][]{Gallerani:2006MNRAS,Pallottini:2014cgmh}:
\be\label{eq_tau}
\tau_{\nu}= f_{\nu_{0}}\left(\pi e^{2} /m_{e} \nu_{0}\right) N_{\rm CII}\psi\, ,
\ee
where $f_{\nu_{0}}$ is the oscillator strength of the \CII~transition \citep[e.g.][]{Basu:2004na}, $e$ and $m_{e}$ are the electron charge and mass, $N_{\rm CII}$ is the \CII~column density\footnote{Note within our formalism the ISM does not contribute to $\tau_{\nu}$ in eq. \ref{eq_tau}, since ISM emission is accounted via the \citetalias{valliniSUB} model (see Sec. \ref{sec_gal_emission}).}, $\psi=\psi((\nu-\nu_{0})/\Delta\nu_{D})$ is the normalized line profile \citep[e.g.][]{Meiksin:2009RvMP}, $\Delta\nu_{D}=(\nu_{0}/c)\sqrt{2k_{B}T_{k}/m_{\rm C}}$ is the thermal Doppler broadening, $k_{B}$ the Boltzmann constant, $m_{\rm C}$ the carbon atom mass, and $c$ the speed of light. We specify that we are focusing on CMB fluctuations produced by \CIIion~ions, but the same formalism can be extended to other fine structure lines.

In Fig. \ref{fig_esempio} we show an example of \CII-induced CMB distortions/fluctuations\footnote{In the CMB terminology, \emph{distortions} indicate spectral variations, while \emph{anisotropies} refer to spatial variations. In this work we study spectral distortions that are induced by a non-uniform distribution of metals. Throughout the paper, we refer to them as \emph{fluctuations}.} for a single line of sight (l.o.s.) extracted from a simulation snapshot centered at $z=6$. In the left panel we plot $\diinu$ as a function of frequency ($\nu$, lower axis) and velocity ($v$, upper axis); in the right panel, the metal patch responsible for the fluctuations is depicted by plotting the relevant physical characteristics (from top to bottom $\Delta$, $Z$ and $v_p$) as a function of comoving distance (${\rm kpc}\,h^{-1}$) along the simulated l.o.s.. The origins of the velocity (left panel) and distance (right panel) axis are shifted for displaying purpose.

Absorption ($\diinu<0$) and emission ($\diinu>0$) features are highlighted with a filled red and blue region, respectively. At $\nu\simeq271.78$ GHz we can see a single absorption peak (labelled $A_{1}$ in the Figure); the emission is instead characterized by a double peaked structure: $E_{1}$ at $\nu\simeq271.81$ GHz and $E_{2}$ at $\nu\simeq271.85$ GHz. Both $A_{1}$ and $E_{1}$ have FWHMs of order $\simeq20\, {\rm km}\,{\rm s}^{-1}$, while the FWHM of $E_{2}$ is larger, $\simeq40\, {\rm km}\,{\rm s}^{-1}$. The l.o.s. intersects the metal patch for $\simeq 0.2\, {\rm Mpc}\,h^{-1}$. The CMB interacts with a CGM peak ($\Delta\simeq10^{2}$) at $x\simeq-30\,{\rm kpc}\,h^{-1}$ and an IGM peak ($\Delta\simeq10^{0.5}$) at $x\simeq+75\,{\rm kpc}\,h^{-1}$.

The CGM is responsible for both the absorption $A_{1}$ and the emission $E_{1}$ features. The change from absorption to emission is caused by the change of sign in $v_p$. As $N_{\rm CII}\propto Z\Delta$, the maximum of the CGM signal is produced rightwards of the $\Delta$ maximum, i.e. at $x\simeq-5\,{\rm kpc}\,h^{-1}$. At this location $v_p<0$, thus the CGM absorption feature ($A_{1}$) is stronger than the emission ($E_{1}$). The broader emission peak $E_{2}$ is caused by the IGM. Because of the lower density, this emission is one order of magnitude lower than the CGM absorption, and of the same order of the emission $E_{1}$.

%$\Delta I_{\nu}(E_{2})\simeq \Delta I_{\nu}(E_{1}) \ll |\Delta I_{\nu}|(A_{1})$.
The cumulative effect of the patch can be calculated by integrating over the shown bandwidth ($\Delta\nu=1$~GHz). This yields a net absorption, with intensity $\diiDnu\simeq-7\times10^{-7}$. The order of magnitude of the cumulative effect is comparable with that of the amplitude of the fluctuations due to up-scattering of CMB photons with hot electrons ($T>10^4$ K) produced by first stars/galaxies radiation, supernova feedback and structure formation shocks at redshifts $z>10-20$ \citep[e.g.][]{Chluba:2014arXiv}.

\section{CMB fluctuations maps}\label{sec_cmb_th_anal}

\begin{figure*}
\centering
\includegraphics[width=0.49\textwidth]{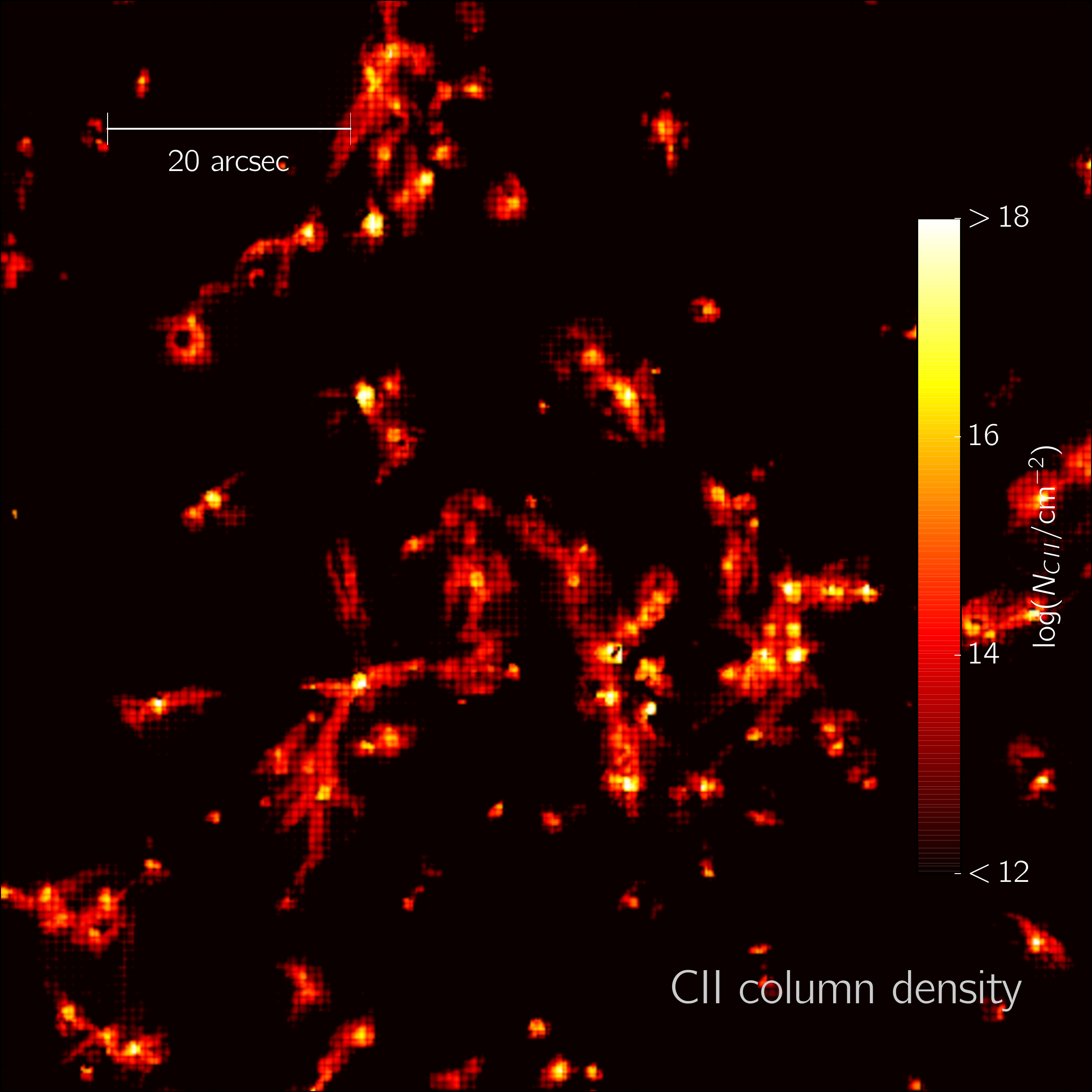}
\includegraphics[width=0.49\textwidth]{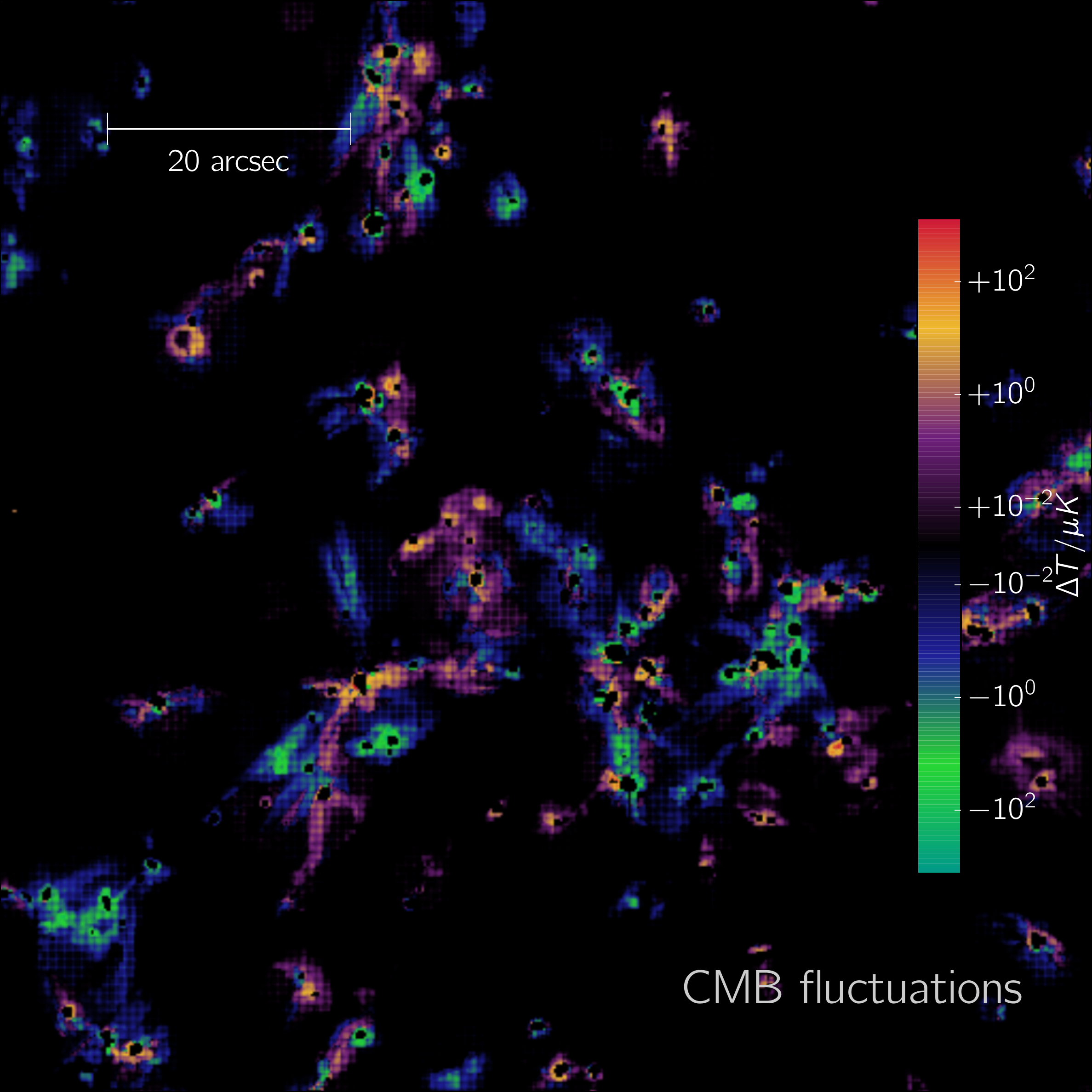}
\caption{\CII~column density map (left panel) and corresponding CMB fluctuations map (right panel) from a simulated field of view (FOV) of $\simeq(90\arcsec)^{2}$ at $z=6$. {\bf Left panel}: map of the \CII~column density ($N_{\rm CII}$) responsible for the CMB fluctuations. The column density is shown for the range $12<\log(N_{\rm CII}/{\rm cm}^{2})<18$ for displaying purposes. {\bf Right panel:} the \CII~CMB fluctuations map is shown as the brightness temperature $\Delta T\equiv\Delta I c^{2}/(2\nu^{2} k_{B})$ integrated on the simulation maximum available bandwidth, i.e. $\Delta\nu\simeq 2.6$~GHz. Note that the black spots inside the fluctuation patches ($|\Delta T|>0$) are caused by the ISM masking. In both panels, the angular scale is indicated as an inset.
\label{fig_mappe_cii_double}}
\end{figure*}

We select a sample of $\simeq4\times 10^{6}$ l.o.s. extracted from a snapshot of the simulation at $z=6$ to compute maps of CMB metal fluctuations. At this redshift, the \citetalias{Pallottini:2014_sim} simulation is characterized by a maximum field of view (FOV) of $\simeq(350\arcsec)^{2}$ and an adaptive spatial resolution down to $\simeq 500\,{\rm pc}$, implying that maps are resolved at angular scales of $\theta_{\rm res}\simeq0.2\arcsec$.

Fig. \ref{fig_mappe_cii_double} shows the \CII~column density map (left panel) and the corresponding CMB fluctuation map (right panel) for a ${\rm FOV}\simeq(90\arcsec)^{2}$. The \CII~column densities are due to projection of the metal bubbles that originate around galaxies, pollute the IGM/CGM and extend in the cosmic web filaments. The density profiles around galaxies are self-similar once scaled with the virial radius of the parent dark matter halo \citep[][]{Pallottini:2014cgmh}, and metals usually extend out to $\simeq 10\, r_{\rm vir}$. Thus, extended patches with high $N_{\rm CII}$ values are found nearby older (more massive) galaxies, which have more time to increase the metallicity of the central CGM part and pollute the surrounding IGM environment. As a reference, the ISM typically has $\log(N_{\rm CII}/{\rm cm}^{-2})\gsim18$, the CGM (IGM) can be enriched up to $\log(N_{\rm CII}/{\rm cm}^{-2})\gsim16$ ($\log(N_{\rm CII}/{\rm cm}^{-2})\gsim12$).

In the right panel of Fig. \ref{fig_mappe_cii_double}, CMB fluctuations are expressed in terms of differential brightness temperature, $\Delta T\equiv \Delta I c^{2}/(2\nu^{2} k_{B})$, where $\Delta I$ is obtained by integrating $\Delta I_{\nu}$ (see eq. \ref{eq_maoli}) on the total bandwidth of the simulation, i.e. $\Delta\nu\simeq 2.6$~GHz. The signal ranges from emissions up to $\Delta T\simeq+10^{2}\mu{\rm K}$ down to absorptions of order $\Delta T\simeq-10^{2}\mu{\rm K}$. The signal is negligible ($|\Delta T |\lsim 10^{-3}\mu{\rm K}$) in $\simeq60\%$ of the selected FOV, consistently with the analysis of the metal filling factor at this redshift \citep[][]{Pallottini:2014_sim}.

The morphology of CMB fluctuations follows the $N_{\rm CII}$ distribution. The lack of an exact match between the respective maps is due to the dependence of $\Delta T$ from the peculiar velocity field. Absorption and emission features can arise from the differential velocity structure inside a single metal bubble. This is the spatial analog of the emission/absorption features as a function of $\nu$ for the l.o.s. shown in Fig. \ref{fig_esempio}.

Fluctuations are preferentially located in correspondence of $\log (N_{\rm CII}/{\rm cm}^{-2})>12$ patches, characterized by typical sizes $\theta\simeq20\arcsec$; local maxima of the fluctuations ($|\Delta T|>10^{2}\mu{\rm K}$) are usually found in smaller spots ($\theta\simeq1\arcsec$) characterized by $\log (N_{\rm CII}/{\rm cm}^{-2})\simeq 17$. This result is confirmed through the analysis of the CMB fluctuations power spectrum that peaks at angular scale $\theta\simeq1\arcsec$, as detailed in App. \ref{sec_app_ps}.

CMB fluctuations on large ($\simeq10\arcsec$) and small ($\simeq1\arcsec$) scales arise from the enriched IGM and CGM, respectively. Smaller scales correspond to the ISM of galaxies. These regions have been masked\footnote{The ISM masking interests less than $0.5\%$ of the total FOV.}, since they do not contribute to CMB fluctuations, while possibly shining as \CII~emitters.

The \CII~is a biased tracer of the cosmic web as it is mostly concentrated in the intersection of the filaments (\quotes{knots} hosting galactic environments) where the density is sufficiently large ($\Delta\gsim10^2$) that recombinations overcome photoionization from the UV background (and possibly local) radiation field. In addition, as several observational and theoretical works have shown, the filling factor of metals at these high-$z$ is relatively small, of the order of $10\%$-$20\%$. So independently of the ionization state, it would be anyway difficult to find carbon in regions of moderate to low density ($\Delta\lsim 5$) corresponding to the filaments.

\section{Mock observations}\label{sez_mock}

\begin{figure*}
\centering
\includegraphics[width=0.49\textwidth]{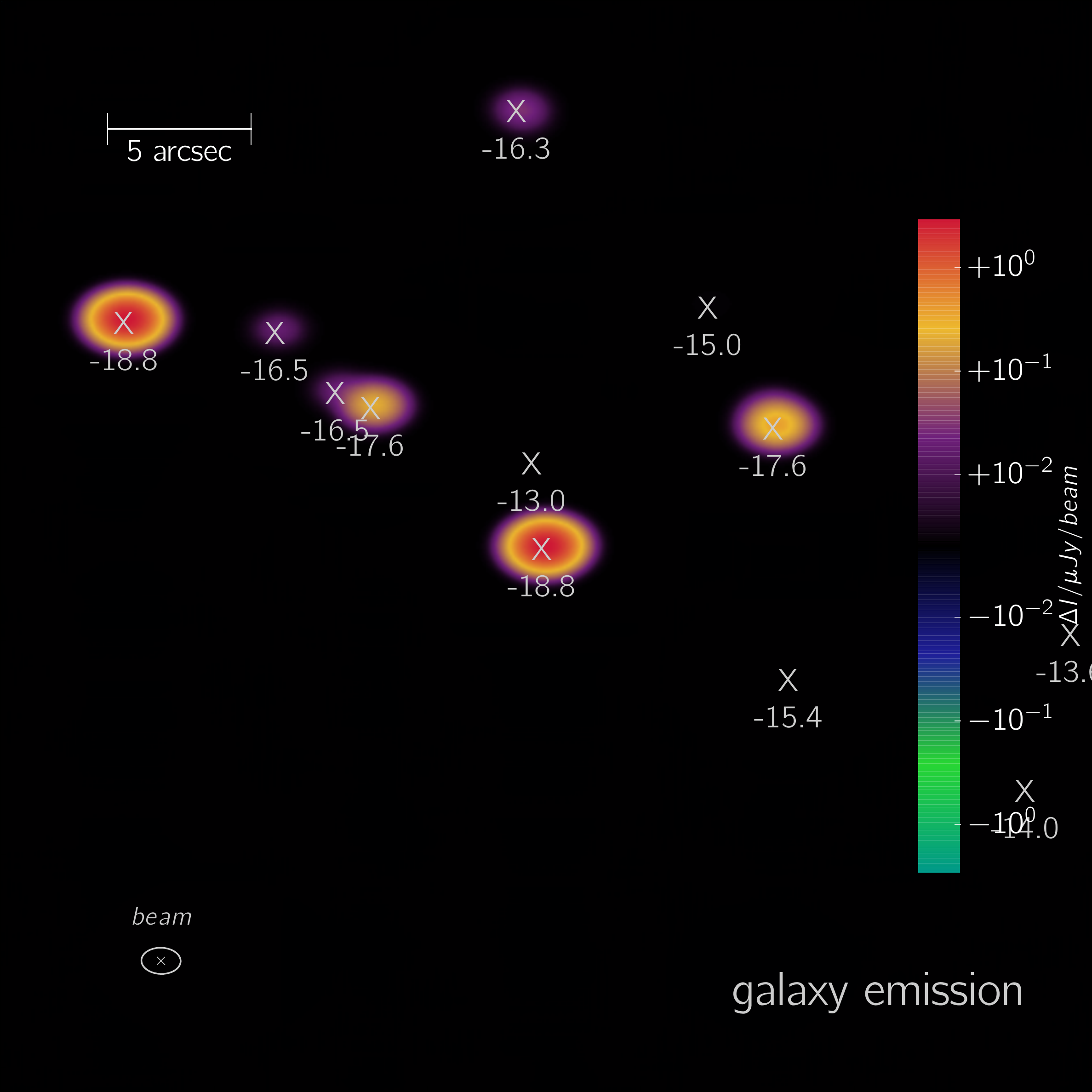}
\includegraphics[width=0.49\textwidth]{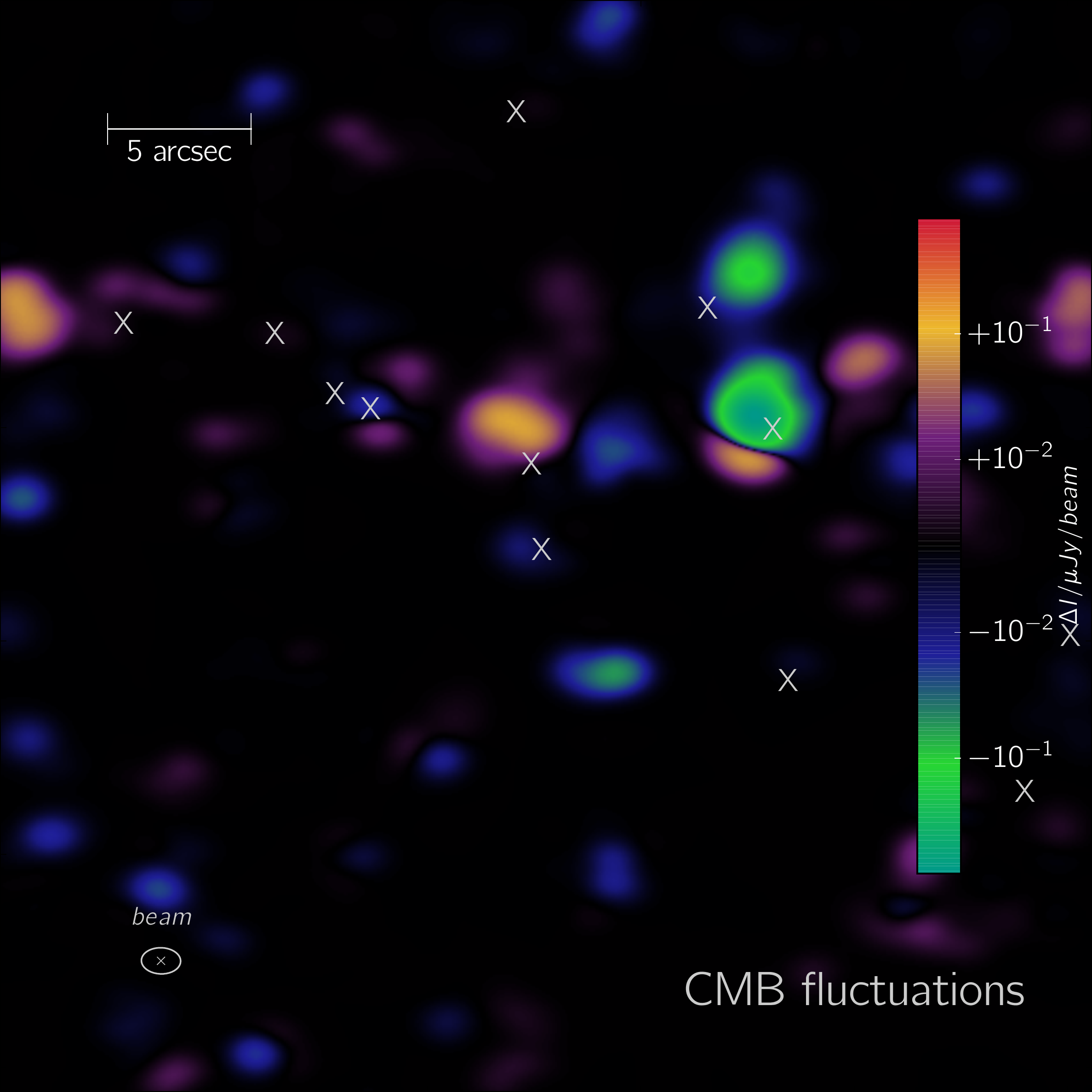}\\
\vspace{2.5pt}
\includegraphics[width=0.49\textwidth]{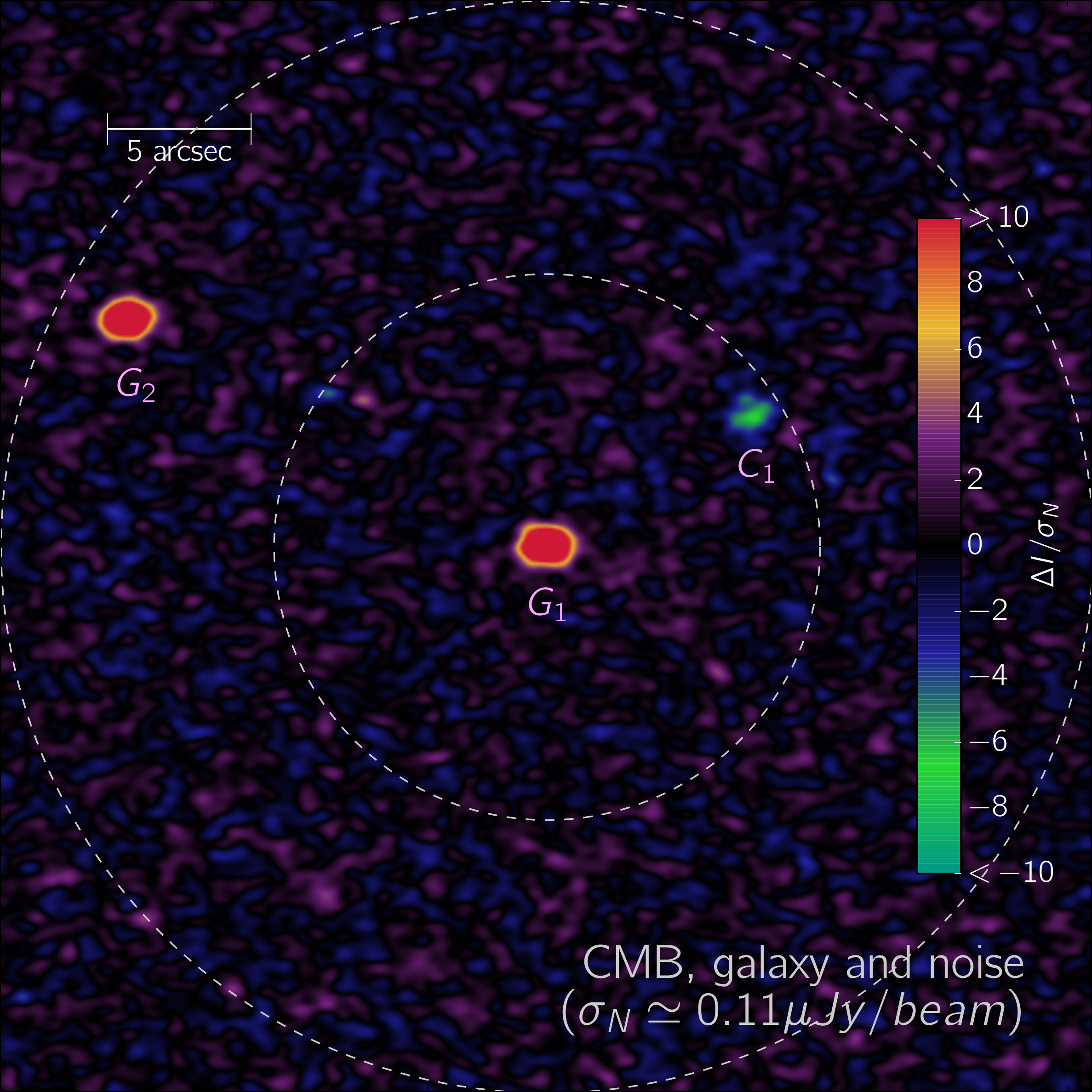}
\includegraphics[width=0.49\textwidth]{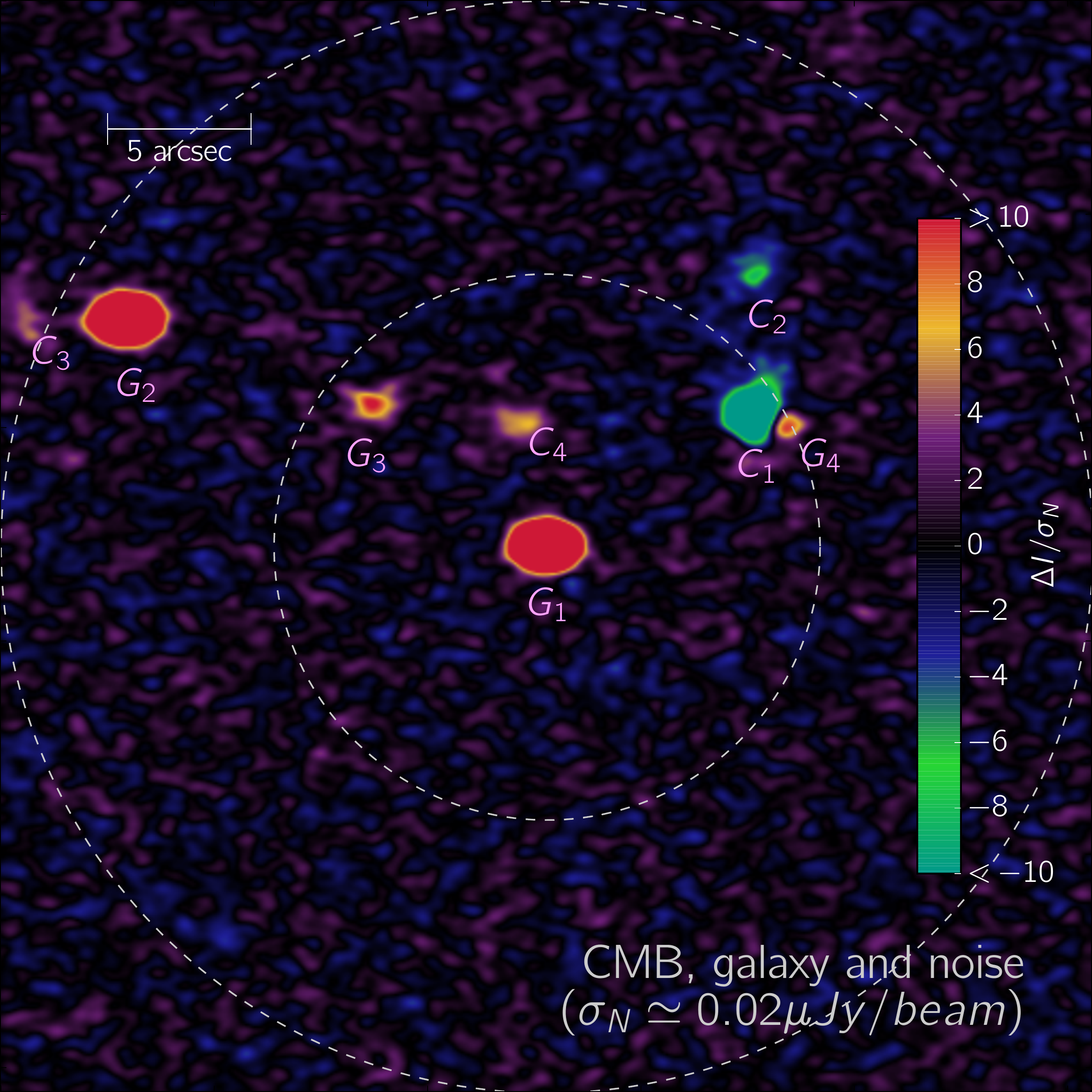}
\caption{Mock ALMA \CII~continuum ($\Delta\nu=8~{\rm GHz}$) map observed with BAND6 with a synthesized beam of $\simeq(1\arcsec)^{2}$.
The maps are shown for a total FOV of $(38.04\simeq as)^2$, twice BAND6 primary beam. In each panel, the map, the angular scale and the synthesized beam are indicated as insets. In the upper panels we separately plot the contribution for galaxy emission and CMB fluctuations. In the lower panels two mock observations (galaxy emission and CMB fluctuations and noise) for two different noise levels are shown.
{\bf Upper left panel}: \CII~signal from galaxy emission. The positions of galaxies are indicated with an `X' and labelled with the corresponding UV magnitude ($M_{\rm UV}$, details in the text). The intensity is expressed as $\mujb$ and, to better appreciate the presence/absence of the signal, the intensity is in symmetric logarithmic scale.
{\bf Upper right panel}: \CII~signal from CMB fluctuations from the same field of view. As for emission from galactic ISM, the intensity scale is nonlinear.
{\bf Lower left panel}: mock observation, obtained by summing the contributions from CMB fluctuation, galaxy emission and noise. The noise distributions fit a Gaussian with r.m.s. $\sigma_{N}\simeq0.11 \mujb$. To augment the readability of the map, the intensity is rescaled with $\sigma_{N}$ and the scale is limited to $|\Delta I|<10\,\sigma_{N}$. To help the description in the text, detections of CMB fluctuations and galaxy emission are indicated with $C_i$ and $G_i$, respectively. For reference, we plot the ALMA primary beam and twice the primary beam with dashed lines.
{\bf Lower right panel}: as in the lower left panel but with $\sigma_{N}\simeq0.02 \mujb$.
\label{fig_mappe_deep}
}
\end{figure*}

Starting from our model for \CII-induced CMB fluctuations and \CII~emission from $z=6$ galaxies, we generate mock observations specifically suited for comparison with ALMA data. The \CII~line at 157.74~$\mu {\rm m}$ from $z=6$ is redshifted in ALMA BAND6 ($1.1-1.4\,{\rm mm}$). At this wavelengths the FOV (primary beam) of ALMA antennas is $\simeq20\arcsec$, and the angular resolution (synthesized beam) ranges from $1.8\arcsec$ to $0.14\arcsec$, for the most compact and the most extended configuration, respectively \citep[e.g.][]{Maiolino:2008NewAR}\footnote{For further details, please refer to the ALMA handbook \url{https://almascience.eso.org/documents-and-tools/cycle-0/alma-technical-handbook/at_download/file}}.

We account for ALMA angular resolution by convolving our maps with the synthesized beam, that can be characterized as a 2D-Gaussian. In this Sec., we assume a major axis of $1.4\arcsec$, a minor axis of $0.9\arcsec$, and a position angle of $91.6~{\rm deg}$. 

We note that the synthesized beam that optimizes the intensity of the signal depends on the experiment that one would like to carry on. In the case of CMB fluctuations, the synthesized beam that optimizes the chance of detection is $\simeq1\arcsec$, since the power spectrum peaks at this scale, as shown in Sec. \ref{sec_cmb_th_anal} and by the power spectrum analysis, reported in the App \ref{sec_app_ps}. Viceversa, for detecting \CII~emission from galaxies, a synthesized beam of $\sim0.5\arcsec$ is more appropriate since it corresponds to typical galaxy sizes at $z\simeq6$, thus minimizing the surface-brightness bias for galaxy detection \citep[e.g.][]{Grazian:2011A&A}. This different observational set-up is further described in Sec. \ref{sec_strategy}.

CMB fluctuations and galaxy emission maps are generated by integrating on $\Delta\nu=8~{\rm GHz}$, i.e. the maximum bandwidth available for continuum observations. This is done in order to maximize both the sensitivity and the possibility of finding CMB signal in the selected FOV. The \citetalias{Pallottini:2014_sim} simulation size limits the maximum available bandwidth to $\Delta\nu=2.6~{\rm GHz}$ (see Sec. \ref{sec_cmb_th_anal}). We generate a longer light-cone by replicating the $z=6$ simulation box. To avoid spurious periodicity effects, we randomize the simulation box by applying translations, rotations and reflections \citep[e.g.][]{Blaizot:2005MNRAS}. Then, the signal from the light-cone is integrated over $\Delta\nu=8~{\rm GHz}$, by taking into account that ALMA continuum band is not contiguous, i.e. it is composed of two $4~{\rm GHz}$ bands, separated by $10~{\rm GHz}$.

Finally, mock maps are combined with noise maps generated with \textlcsc{ALMAOST} \citep[][]{ALMAOST:2011arXiv}. Within the primary beam, the noise map has a Gaussian distribution with zero mean and r.m.s. $\sigma_{N}$; the r.m.s. accounts for instrumental uncertainties and atmospheric conditions. We assume $0.913\,{\rm mm}$ of water vapor for our fiducial observational setting. Using the ALMA full array, we have a continuum sensitivity of $\sigma_{N}\simeq 11.09\,(t/{\rm hr})^{-1/2} (\Delta\nu/8~{\rm GHz})^{-1/2}[\,\mujb]$ for an observational time $t$ at $\nu_{\rm obs}\simeq271~{\rm GHz}$.

In Fig. \ref{fig_mappe_deep} we plot mock maps centered on the same $M_{\rm UV}\simeq-19$ galaxy. In the upper panels, we plot separately the galaxy emission (left) and CMB fluctuations (right) signals, expressed in $\mujb$. The angular scale and the synthesized beam are indicated in the inset. In the galaxy emission map, the symbol `X' indicates locations of galaxies, which are labelled according to their $M_{\rm UV}$. Galaxies are mostly found clustered in the upper region, and the plot shows a clear correlation between galaxy UV magnitude and \CII~emission (see the right panel of Fig. \ref{fig_luminosity} in Sec. \ref{sec_gal_emission}). In fact, brighter \CII~emitters ($\Delta I\simeq \mujb$) correspond to brighter UV galaxies ($M_{\rm UV}\simeq-19$), characterized by a total \CII~flux of the order of $\Delta I\simeq 100~\mu$Jy; dimmer galaxies ($M_{\rm UV}\simeq-16.5$) are identified through \CII~emission peaks of the order of $\Delta I\simeq 10^{-1}\mujb$; the faintest galaxies in the map ($M_{\rm UV}\simeq-15$) show very low \CII~emission, $\Delta I\lsim 10^{-2}\mujb$.

The CMB fluctuations map (upper left panel of Fig. \ref{fig_mappe_deep}) shows various peaks both in absorption and emission. The signal due to CMB fluctuations result to be more extended and offset with respect to galaxy emission (indicated as `X'). This is expected from our analysis of $N_{\rm CII}$ and CMB fluctuations maps (see Fig. \ref{fig_mappe_cii_double}). Typically, emission/absorption features range between $10^{-2}\lsim|\Delta I|/(\mujb)\lsim 10^{-1}$, with local intensity maxima comparable to the emission signal of a $M_{UV}\simeq-17$ galaxy. The strongest peak occurs in absorption and it characterized by $\Delta I\lsim -10^{-1}\mujb$.

In the lower panels of Fig. \ref{fig_mappe_deep}, we plot the total signal, i.e. including the contribution from CMB fluctuations, galactic emission and two different noise levels: $\sigma_{N}\simeq 0.1\mujb$ (left) and $\sigma_{N}\simeq0.02\mujb$ (right) that correspond to $t\simeq 1.3\times10^{4} {\rm hr}$ and $t\simeq 3.1\times10^{5} {\rm hr}$ observing time, respectively. In these panels, the intensity of the signal is rescaled to the noise r.m.s.; additionally, detections of CMB fluctuations and galaxy emission are indicated with $C_i$ and $G_i$, respectively.

For $\sigma_{N}\simeq0.11 \mujb$, the noise covers most of the CMB fluctuations and all the faint peaks of the galactic emission map ($M_{UV}\gsim-17$). Detectable signals arise from the two $M_{UV}\simeq-19$ galaxies ($G_1$ and $G_2$) and from the peak of the CMB fluctuations ($C_1$); the galaxies show $>10\sigma_{N}$ emission peaks, and the fluctuation appears as a $\simeq7\sigma_{N}$ absorption signal.

In the case of the deeper mock observation, for $\sigma_{N}\simeq0.02~\mujb$, $M_{\rm UV}\simeq-17.6$ galaxies are detectable ($G_3$ and $G_4$) at $>6\sigma_{N}$ as \CII~emitters within the primary beam. Note that $G_3$ is a $>10\sigma_{N}$ peak, while $G_4$ show an emission $\simeq 6 \sigma_{N}$, as its signal is partially suppressed by the nearby absorption from CMB fluctuations at $C_1$. Three more CMB fluctuations peaks are visible on the map: a $\simeq 6 \sigma_{N}$ absorption arising outside the primary beam ($C_2$), and two extended $\simeq 5\sigma_{N}$ emissions ($C_3$ and $C_4$) located within $2.5\arcsec$ from the $M_{UV}\simeq-19$ galaxies.

Note that CMB fluctuations in emission can be mistaken for signals arising from galaxies. In principle, the origin of the signal can be distinguished by analyzing its spectral properties. Typically, galaxy \CII~emission lines have FWHMs$\gsim 50-100\, {\rm km}\,{\rm s}^{-1}$ \citep[e.g.][]{Vallini:2013MNRAS,Decarli:2014ApJ,Cormier:2015arXiv}, while CMB fluctuations usually show a double-peak structure in the spectra (see Sec. \ref{sec_cmb_model}, in particular Fig. \ref{fig_esempio}), with narrower lines (FWHM$\simeq30-40\, {\rm km}\,{\rm s}^{-1}$) both in absorption and emission.

A more robust discriminant for the nature of the signal can be provided by multiwavelength analysis. By comparing the \CII~map with UV imaging of the field, we can distinguish between the emission from CMB fluctuations and galaxies. Knowing the location and UV luminosity of galaxies in the field, we can infer the expected \CII~flux (see Sec. \ref{sec_gal_emission}, in particular right panel of Fig. \ref{fig_luminosity}). JWST will have a reference limiting magnitude of $M_{\rm UV}\simeq-16$ at $z=6$, thus it seems to be perfectly suited for such comparison.

\section{Observational prospects}\label{sec_strategy}

\begin{figure*}
\centering
\includegraphics[width=0.49\textwidth]{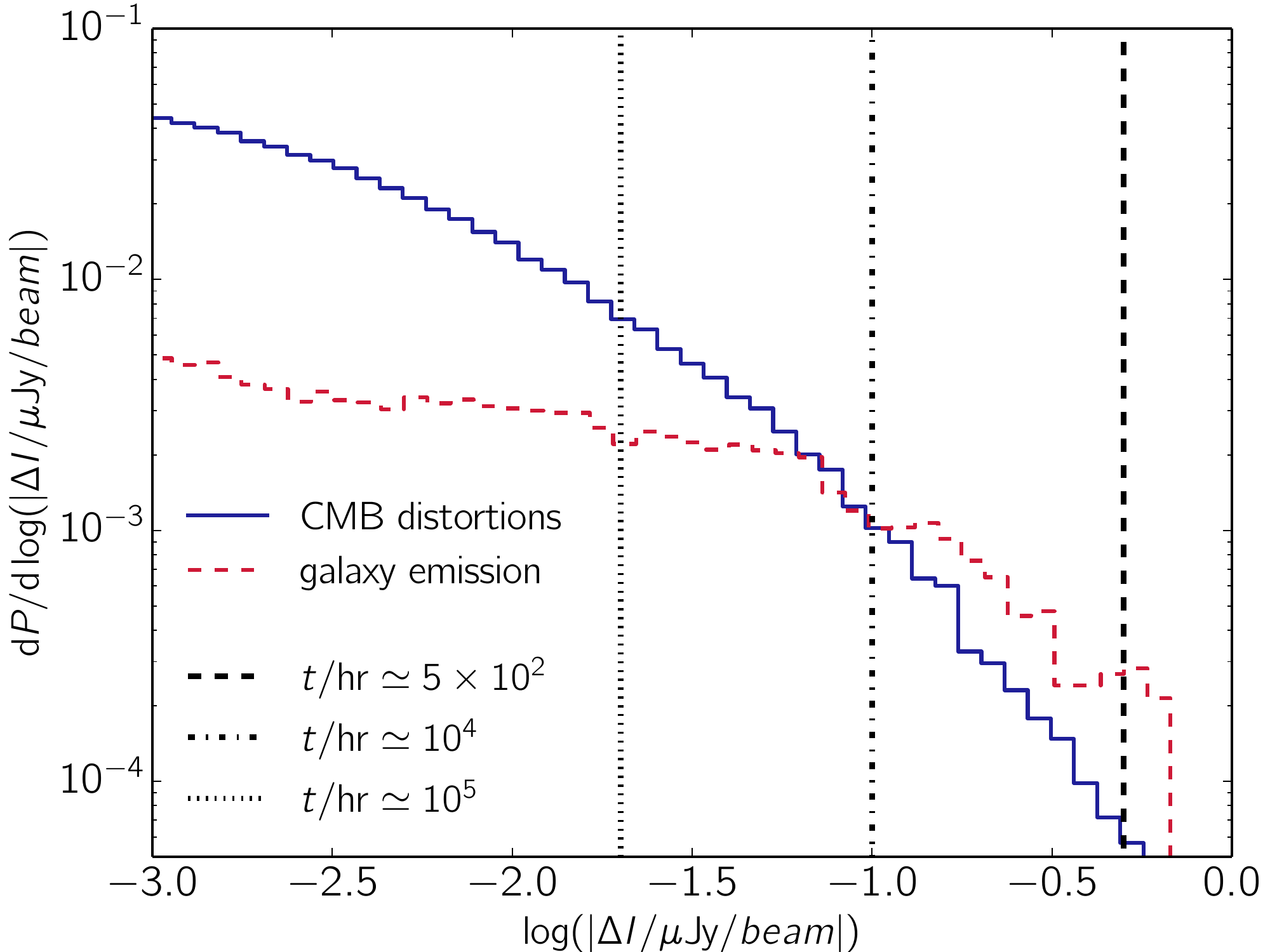}
\includegraphics[width=0.49\textwidth]{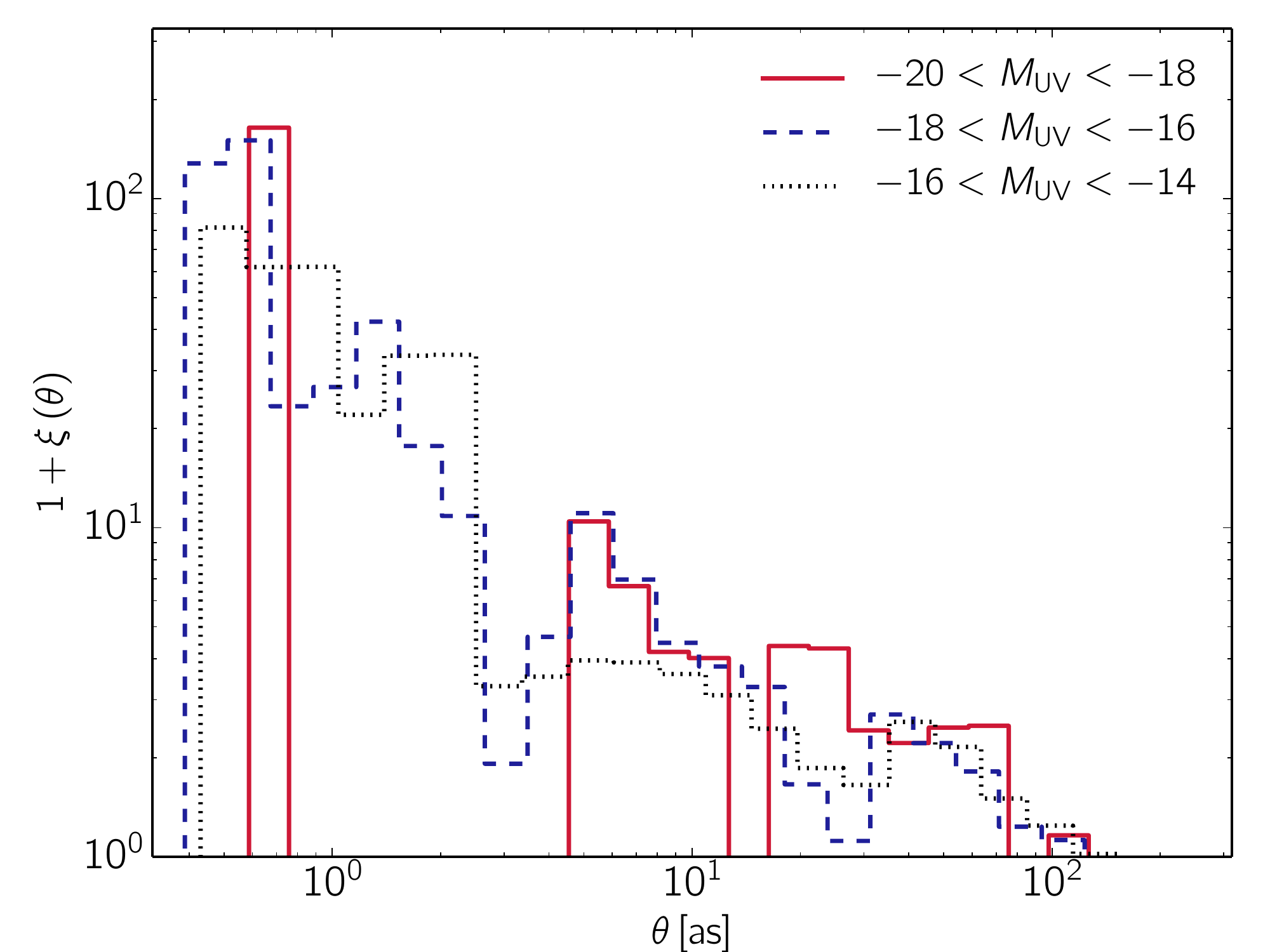}
\caption{
{\bf Left panel}: PDFs of the absolute value of the signal from CMB fluctuations (blue solid line) and galaxy emission (red solid) are plotted. Both signals account for the assumed ALMA synthesized beam (see Sec. \ref{sez_mock}). The PDF integral is normalized to unity, and the plot is cut at $|\Delta I|=10^{-3}\mujb$ for displaying purpose. As a reference, with vertical lines we plot several flux detection limits: the corresponding ALMA observational time is indicated in the legend.
{\bf Right panel}: Angular cross-correlation ($\xi$) function between high CMB fluctuation peaks ($|\Delta I|>0.1\mujb$, see text) and galaxies with limiting magnitude $M_{\rm UV} = -18,~-16~{\rm and}~-14$ (plotted with red, blue and black lines, respectively).
\label{fig_stats}
}
\end{figure*}

The amplitude of \CII-induced CMB fluctuations is expected to be $|\Delta I|\lsim 1\mujb$, as discussed in the previous section. With the aim of defining the optimal observational strategy for detecting such weak signals, we first consider the feasibility of a blind survey experiment with ALMA. We compute the probability distribution functions (PDFs) of the intensity of \CII-induced CMB fluctuations and \CII~galaxy emission at $z=6$. We consider the signals extracted from the whole simulation FOV, $\simeq(350\arcsec)^{2}$, on the $\Delta\nu\simeq 2.6\,{\rm GHz}$ bandwidth centered at the frequency $\nu_{\rm obs}\simeq271~{\rm GHz}$. For both signals, the intensity accounts for the $\simeq (1\arcsec)^{2}$ ALMA synthesized beam introduced in Sec. \ref{sez_mock}. For each PDF, the corresponding integral is normalized to unity.

In the left panel of Fig. \ref{fig_stats}, we plot the PDFs for CMB fluctuations and galaxy emission with blue solid and red dashed lines, respectively. The plot is cut at $|\Delta I|=10^{-3}\mujb$ for displaying purpose. As a reference, with vertical lines we plot 1-$\sigma$ flux detection limits that correspond to the ALMA observing time indicated in the legend, calculated with the ALMA Sensitivity Calculator\footnote{\url{https://almascience.eso.org/proposing/sensitivity-calculator}}. For the CMB fluctuations, the PDF rapidly decreases from $10^{-1.5}$ at $|\Delta I|\simeq 10^{-3}\mujb$ down to $10^{-4}$ at $|\Delta I|\simeq 10^{-0.5}\mujb$. For the galaxy emission, the PDF decrease is less steep, and the two PDFs cross at $|\Delta I|\simeq 10^{-1}\mujb$.

We indicate with $\sigma_{\rm N}$ the 1-$\sigma$ flux detection limit of the ALMA blind survey. Then the integral of the PDF above $\sigma_{\rm N}$ provides the probability $P_{\rm dis}$ and $P_{\rm gal}$ of detecting CMB fluctuations and galaxy emission, respectively. For $\sigma_{\rm N}=10^{-3}\mujb$ (corresponding to an ALMA observing time $t\simeq 10^{6} {\rm hr}$), we find $P_{\rm dis}\simeq 4\%$ and $P_{\rm gal}\simeq 1\%$. Given the current telescope sensitivity, if we consider a more realistic -- thought challenging -- observation, i.e. $\sigma_{N}=10^{-1}\mujb$ ($t\simeq 10^{4} {\rm hr}$), we find $P_{\rm gal}\simeq P_{\rm dis}\simeq0.05\%$. This highlights that it is extremely difficult to detect CMB fluctuations with a blind survey.

Therefore, as an alternative strategy we consider the search for CMB fluctuations in the close proximity of high-$z$ galaxies. We compute the cross-correlation ($\xi$) between CMB fluctuations peaks and UV galaxy emission. We calculate $\xi$ by using an extension of the Hamilton estimator \citep[e.g.][]{Nollenberg2005ApJ}:
\be
\xi(\theta) +1= (D_{G}D_{C} \langle D_{R}D_{R} \rangle ) \slash (\langle D_{G} D_{R}\rangle \langle D_{C} D_{R}\rangle)\, ,
\ee
where $D_{i}D_{j}$ is the number of data pairs from sets $i$ and $j$ within angular distance $\theta$, the subscripts $G$ and $C$ indicate the set of positions for galaxy and CMB fluctuations peaks respectively, and $R$ labels a set of locations extracted from a random uniform distribution; the operator $\langle\dots\rangle$ indicates the average on different realizations of $R$. Both the dimension of the set $R$ and the number of realizations for the average are fixed to reach a suitable convergence for $\xi$.

For what concerns CMB fluctuations, we consider only strong peaks, i.e. fluctuations with $|\Delta I|>0.1\mujb$. Within the simulated FOV we find $\simeq30$ of such peaks, with signal intensity up to $0.25\mujb$ in emission and down to $-0.82\mujb$ in absorption. For these signals, the mean $|\Delta I|$ is $\simeq 0.23\mujb$ and the r.m.s. is $\simeq0.25\mujb$. For what concerns galaxies, we select sets of objects characterized by different UV limiting magnitudes.

In the right panel of Fig. \ref{fig_stats}, we plot with red, blue and black lines the cross-correlation calculated for galaxies with $M_{\rm UV}< -18,\, -16\, {\rm ~and}\, -14$, respectively. Although there is a clear cross-correlation between strong CMB fluctuations and galaxy emission, we find no strong dependence of the correlation scale from the considered $M_{\rm UV}$ limits\footnote{The cross-correlation scale does not strongly depend on the considered $M_{\rm UV}$ limits because of the following. Brighter (more massive) galaxies are expected to be surrounded by more extended and enriched CGM. However CMB fluctuations linearly depend on the velocity field, thus there is a lack of correspondence between high $N_{\rm CII}$ values and strong fluctuations (see Sec \ref{sec_cmb_th_anal} for details, in particular Fig. \ref{fig_mappe_cii_double}).}. Therefore, the most promising -- though challenging -- observational strategy for detecting \CII-induced CMB fluctuations is to perform a deep ($\sigma_{N}\lsim0.1\mujb$) ALMA BAND6 observation pointing at a $M_{\rm UV}\simeq-19$ known galaxy, e.g. in the HDF south, where the deepest photometry is available. This observational strategy is detailed in Sec. \ref{sez_mock}.

From Fig. \ref{fig_stats} it appears that \CII~emission detection is very challenging even from the bright end galaxies in our simulation ($M_{\rm UV}\gsim-19$). As anticipated in Sec. \ref{sez_mock}, this is critically dependent on the assumed $\simeq1\arcsec$ synthesized beam. A galaxy at $M_{\rm UV}\simeq-19$ has a \CII~emission of $\simeq 100\mu{\rm Jy}$ (see Fig. \ref{fig_luminosity} and eq. \ref{eq_flux_fit}). This correspond to a flux of $\simeq 1 \mujb$ in the ALMA map with a $\simeq1\arcsec$ synthesized beam. Hence, with this set-up, $\simeq10^{3}$ hours of integration time are required in order to detect the signal at $\simeq4\sigma$.

The detectability drastically improves when using instead a synthesized beam similar to the galaxy size, i.e. $\simeq0.5\arcsec$, like the one used in recent high-$z$ observations \citep{Maiolino:2015arXiv,capak:2015arXiv,Willott:2015arXiv15}. With this set-up, by using eq. \ref{eq_flux_fit} and the ALMA Sensitivity Calculator, we find that a $M_{\rm UV} = -18,-19,-20$ galaxy can be detected at $4\sigma$ in $\simeq 2000, 40, 1$ hours, respectively. Indeed, \CII~has been detected in high-$z$ galaxies with magnitudes up to $M_{\rm UV}> -20$ with $t\sim 1\,{\rm hr}$ \citep{Maiolino:2015arXiv,Willott:2015arXiv15}. With a more demanding observational time of $\sim 40\,{\rm hr}$, we would be able to detect $M_{\rm UV}\simeq -19$ galaxies at $z\simeq6$, thus sampling galaxies more similar to the true reionization sources.

As noted in \citet{pallottini_cr7:2015}, at $z=6$ \citetalias{Pallottini:2014_sim} galaxies have a specific ${\rm SFR}$ of ${\rm sSFR}={SFR}/M_{\star}\simeq2.5\, {\rm Gyr}^{-1}$, that is consistent with observations by \citet[][]{Daddi:2007ApJ} and \citet{Gonzalez:2011}. However, more recent observations seem to favor an higher ${\rm sSFR}$ \citep[e.g.][ ${\rm sSFR}=5\,{\rm Gyr}^{-1}$]{Stark:2013ApJ}. The latter ${\rm sSFR}$ would yield a \CII~emission higher by a factor $\simeq 2$, namely a required observing time a factor $\simeq 0.7$ smaller. An increase of ${\rm sSFR}$ results in an increase of metal production, that may enhance the signal of CMB fluctuations by up to a factor $\sim2$. However, it is difficult to quantify this enhancement, because of the complexity of the metal enrichment process.

\section{Conclusions}\label{sec_conclusions}

We have studied the possibility of mapping heavy elements via far infrared (FIR) emission from the interstellar medium (ISM) of high-$z$ galaxies and cosmic microwave background (CMB) fluctuations induced by metals in the intergalactic medium (IGM). We focus on high-$z$ low mass ($M_{\star}\lsim 10^{10}\msun$) galaxies, which are expected to be abundant at $z\sim6$, and to represent the first ($z\gsim10$) efficient metal polluters of the IGM \citep[e.g.][]{Madau:2001ApJ,Ferrara:2008IAUS} and sources of reionization \citep[e.g.][]{Barkana:2001PhR}.

Among the FIR emission lines, the \CII~{\small$\left(^{2}P_{3/2} \rightarrow\,^{2}P_{1/2}\right)$} transition at 157.74~$\mu m$ is the brightest \citep[e.g.][]{Crawford:1985ApJ,Madden:1997ApJ}, and it is used to trace and characterize $z\sim 6$ galaxies \citep[e.g.][]{Carilli:2013ARA&A}. We compute both the \CII~emission arising from the ISM of $z=6$ galaxies and the amplitude of \CII-induced\footnote{The presented method can be straightforwardly extended to other FIR lines, as \NII~at 122~$\mu {\rm m}$, \OIII~at 88~$\mu {\rm m}$, [\SiI]~at 129~$\mu {\rm m}$ and \OI~at 63~$\mu {\rm m}$.} CMB fluctuations from metals in the IGM and circumgalactic medium (CGM).

We use state-of-the-art high-$z$ hydrodynamical simulations \citep{Pallottini:2014_sim} that follow the evolution of the cosmic metal enrichment from $z=10$ from $z=4$. We calculate the galactic \CII~signal by using an ISM model \citep{Vallini:2013MNRAS,valliniSUB} that accounts for the detailed sub-kpc scales structure of the emission, namely for the cold and warm neutral medium and in high density photodissociation regions (PDRs). We calculate the CMB fluctuations induced by \CIIion~ions by modelling the resonant scattering between the CMB and metals in the IGM/CGM \citep[e.g.][]{Maoli:1996ApJ,Basu:2004na} and we predict the related emission and absorption features arising in the FIR band.

We then carefully analyze the theoretical signal of CMB fluctuations from the IGM/CGM. While the effects of resonant scattering of CMB photons by very early metals ($z\gg10$) were studied with analytical models and simple assumptions about the metal distribution \citep[e.g.][]{Basu:2004na,Schleicher:2008AA}, the present study is the first based on hydrodynamical numerical simulations extending to observable epochs ($z\sim6$). In terms of the differential brightness temperature, $\Delta T\equiv \Delta I c^{2}/(2\nu^{2} k_{B})$, we calculate the signal from $z=6$ integrating on the simulation available bandwidth, $\Delta\nu=2.6~{\rm GHz}$. We find that the metal-induced fluctuation signal is in the range $\Delta T\simeq\pm10^{2}\mu{\rm K}$, i.e. can be seen either in emission or absorption. The peak of the signal is found on scales of $\theta\simeq1\arcsec$, in correspondence of CGM absorption systems characterized by \CIIion~column density of $\log(N_{\rm CII}/{\rm cm}^{-2})\simeq16$.

To test the detectability of the \CII~signal, we have constructed and analyzed mock observations specifically suited for comparison with ALMA BAND6 data.

We predict that \CII~emission is correlated with $M_{\rm UV}$. At $M_{\rm UV}<-20$, the faintest high-$z$ galaxy from which \CII~emission is detected, our relation (eq. \ref{eq_flux_fit}) is in good agreement with recent observations \citep[][]{Maiolino:2015arXiv,capak:2015arXiv,Willott:2015arXiv15}. We find that a $M_{\rm UV}= -18,\, -19,\, -20$ galaxy can be detected at $4\sigma$ in $\simeq 2000,\, 40,\, 1$ hours, respectively. Indeed, \CII~has been detected in high-$z$ galaxies with magnitudes up to $M_{\rm UV}> -20$ with $t\sim 1\,{\rm hr}$ \citep{Maiolino:2015arXiv,Willott:2015arXiv15}. With a more demanding observational time of $\sim 40\,{\rm hr}$ at $z\simeq6$ we would be able to detect $M_{\rm UV}\simeq -19$ galaxies, thus sampling galaxies more similar to the true reionization sources. Additionally, the relation is the analogous of the local ($z=0$) $L_{\rm CII}$-SFR relation \citep[][]{DeLooze:2014AA}. Eq. \ref{eq_flux_fit} highlights the possibility of using \CII~as a tracer of star formation activity, that can be used independently from UV flux determination. This is particularly important at high-$z$ as, contrary to UV light, \CII~emission is not affected by dust extinction.

FIR observations allow in principle to simultaneously detect galaxies and to map their surrounding CGM through \CII~induced CMB fluctuations. The detection of this signal would be a breakthrough for our understanding of the early phases of galaxy formation and cosmic enrichment. The CGM is the interface regulating the outflows of enriched material from the galaxy and the inflows of pristine gas from the IGM; it has been probed so far up to $z\sim2$ with absorption line experiments towards background sources, QSO or galaxies \citep[e.g.][]{Steidel:2010ApJ,Churchill:2013arXiv,Liang:2014mnras}. The intervening CGM associated with a foreground galaxy leaves an absorption feature in observed spectra. With a large number of galaxy-absorber pairs, it is then possible to statistically determine the equivalent width of a given absorption line as a function of the line of sight impact parameter. Detecting CMB fluctuations is equivalent to actually map the bidimensional distribution of metals in the CGM, and it would allow to extend the enrichment study at higher redshift.

We have compared the efficiency of different observational strategies for detecting \CII~induced CMB fluctuations. We find that this signal is very faint, e.g. $|\Delta I|\simeq10^{-1}\mujb$. Because of the signal amplitude and the \CII~emission filling factor, we find that a blind experiment yields a low probability for the signal detection. However, in our maps we see strong (i.e. $|\Delta I|\gsim10^{-1}\mujb$) CMB fluctuations, that are characterized by a mean $|\Delta I|$ of $\simeq 0.23\mujb$ a r.m.s. of $\simeq0.25\mujb$ and maximum $|\Delta I|$ up to $0.82\mujb$. On average, these fluctuations have signals comparable to the typical \CII~emission from a $M_{\rm UV}\simeq-18$ galaxy. Our analysis highlights that strong CMB fluctuations are found typically within $\sim10\arcsec$ of galaxies, regardless of their $M_{\rm UV}$.

We have also considered an alternative observational strategy to detect (strong) CMB fluctuations that consists of a deep ALMA pointing in a field where known $M_{\rm UV}\simeq-19$ galaxies (e.g. in the HDF south) are present. Our model predicts that, for an ALMA sensitivity of $\sigma_{N}=0.1\mujb$ CMB fluctuations would be detected at a confidence level $\sigma>3,\,7$ with a probability $\simeq25\%,\,5\%$, respectively, while for $\sigma_{N}=0.05\mujb$ we expect a detection of CMB fluctuations with c.l. $\sigma>3,\,5,\,7$ with a probability $\simeq70\%,\,30\%,\,15\%$, respectively. We note that the sensitivity requested for detecting \CII~induced CMB fluctuations ($\sigma_{N}\simlt 0.1\mujb$) corresponds to extremely long ALMA observing times ($t\simgt 10^{4} {\rm hr}$), making the detection of this signal extremely challenging with current facilities.

However, by stacking deep ALMA observations ($\nu_{\rm obs}\simeq 272$~GHz) of several $M_{\rm UV}\simeq-19$ (lensed) galaxies it may be possible to \quotes{statistically} detect such elusive CMB fluctuations induced by \CIIion~ions at $z\simeq 6$. The same experiment, repeated at different frequencies ($\nu_{\rm obs}\simeq 1900/(1+z)$~GHz) would allow to track the metal enrichment history through cosmic times. We defer this further statistical analysis to a future study.

\section*{Acknowledgments}
We thank A. Mesinger for useful discussions.

\bibliographystyle{mnras}
\bibliography{supp}
\bsp

\appendix
\section{Power spectrum}\label{sec_app_ps}

\begin{figure}
\centering
\includegraphics[width=0.49\textwidth]{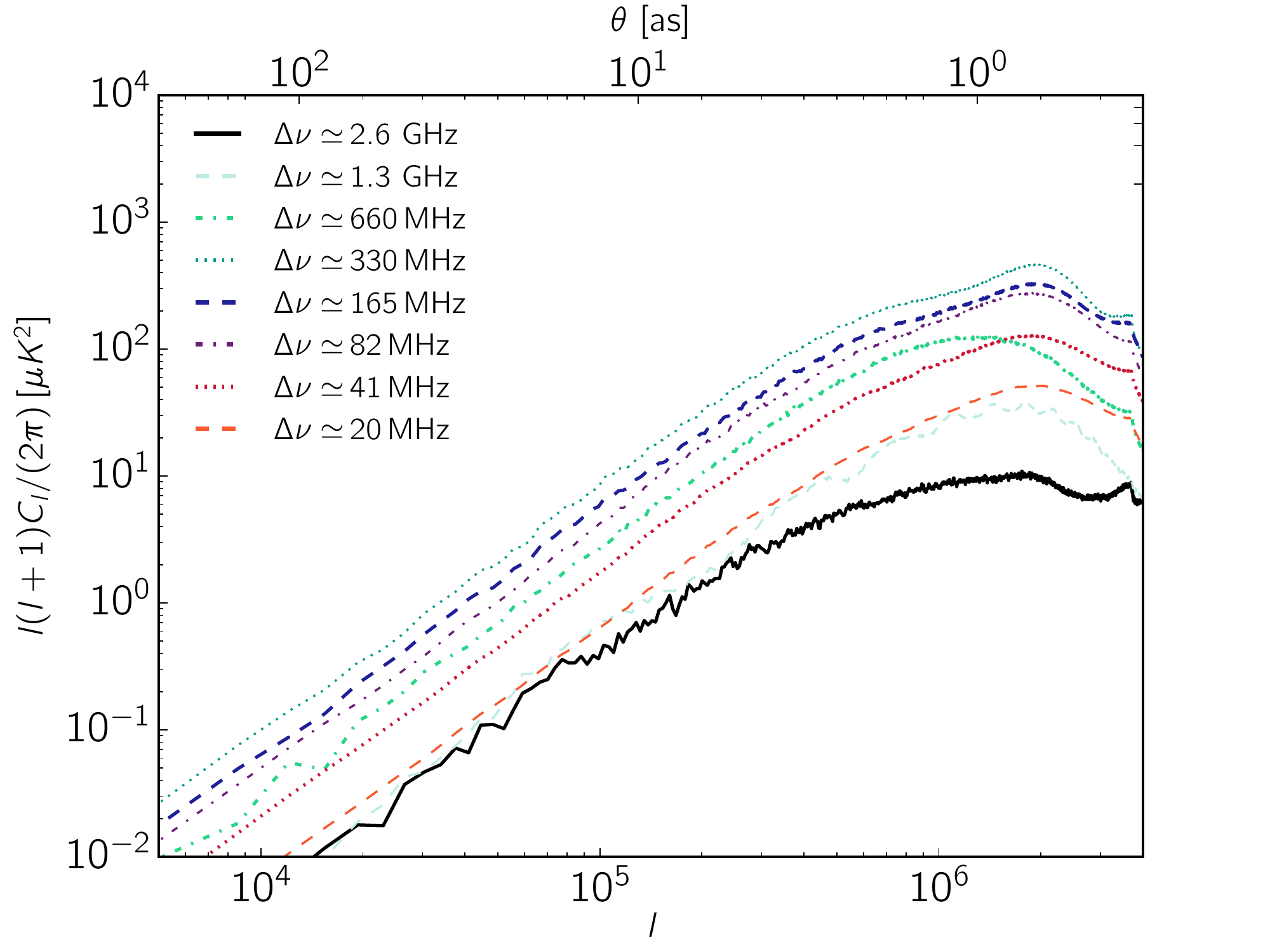}\\
\includegraphics[width=0.49\textwidth]{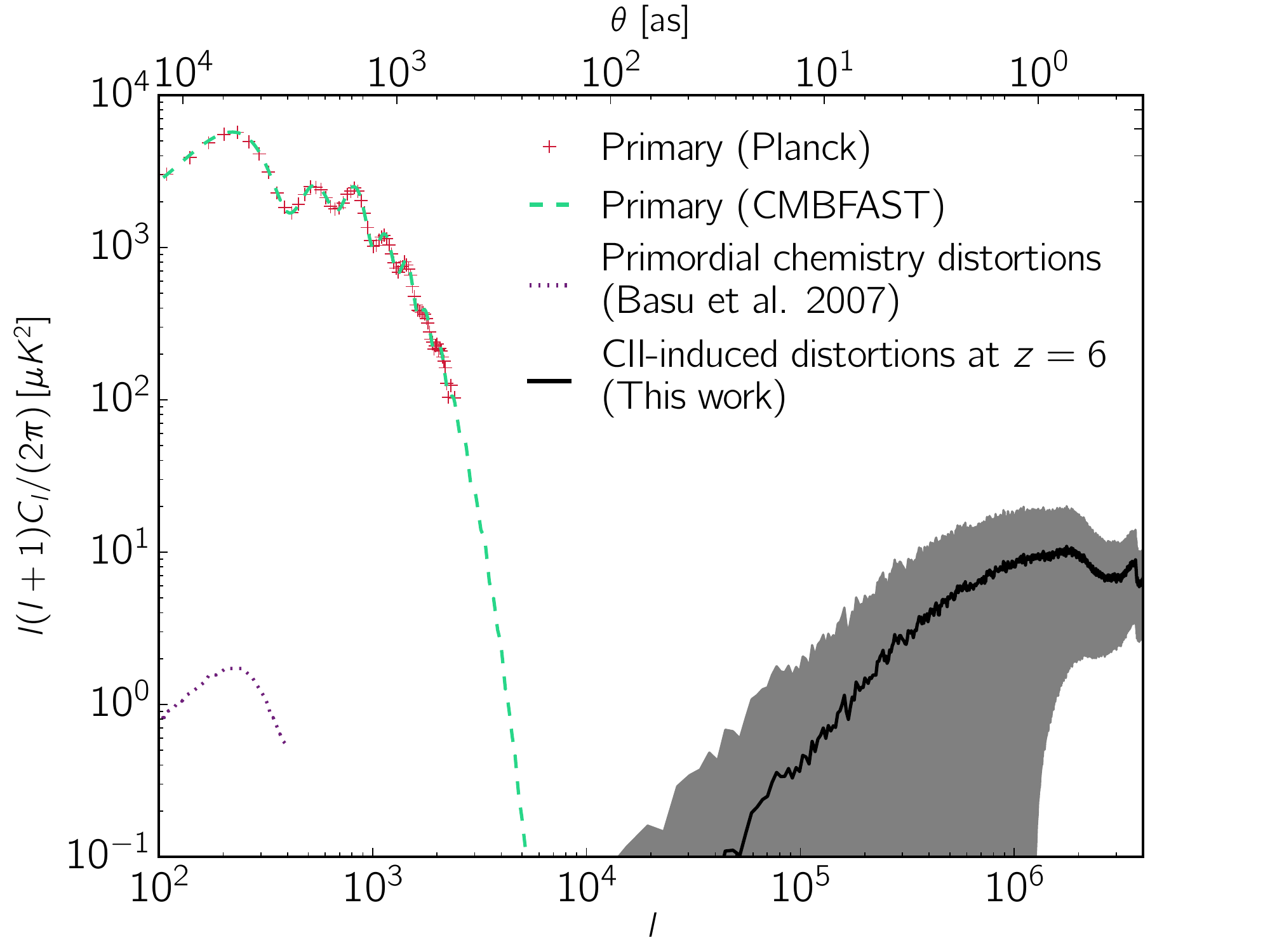}\\
\includegraphics[width=0.49\textwidth]{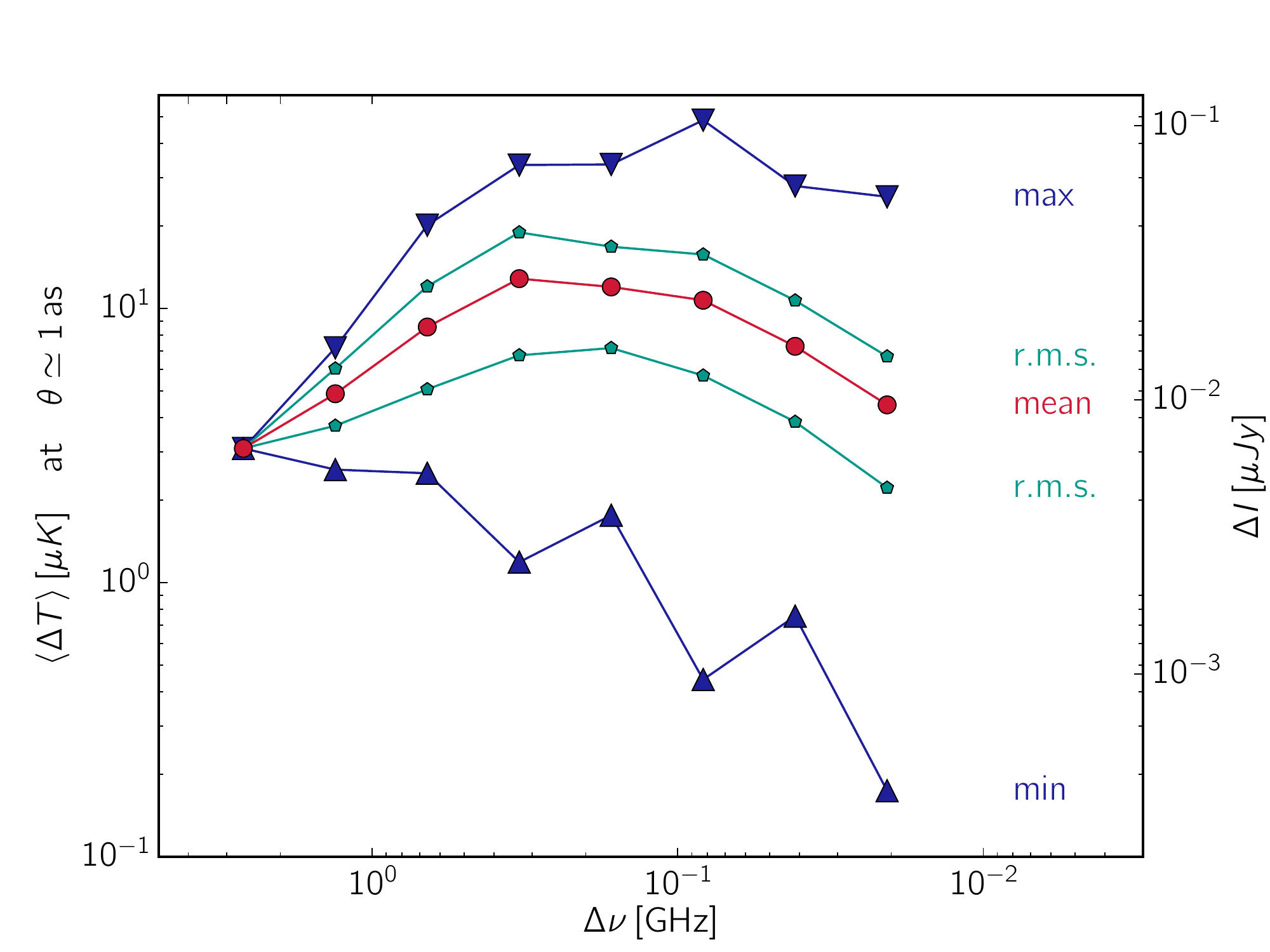}
\caption{
\CII-induced CMB power spectrum (PS, $P_{l}=l\, (l+1)\, C_{l}/2\pi$) as a function of the angular scale, $l$ and $\theta$ in the lower and upper axis, respectively.
In the {\bf upper panel} we plot $P_{l}$ calculated for different bandwidths, as indicated in the legend. In the {\bf central panel} we compare our result with other PSs, as detailed in the legend and in the text. A color version of the plot is available in the online version of the paper.
{\bf Lower panel}: maximum of PS ($\langle\Delta T\rangle=\sqrt{P_{l}}$ at $\theta\simeq1\arcsec$) as a function of the bandwidth ($\Delta\nu$). For each $\Delta\nu$ we plot the mean (red circles), the r.m.s. variance (green pentagons) and the max/min values (blue triangles). The maximum is indicated in unit of $\mu{\rm K}$ (left axis) and $\mu{\rm Jy}$ (right axis).
\label{fig_ps}}
\end{figure}

We investigate the CMB fluctuations morphology by calculating the angular power spectrum (PS, $C_{l}$) of their amplitude. At $z=6$ the simulation FOV is limited to $\simeq(350\arcsec)^{2}$, therefore the minimum angular scale solved is $l\simeq10^{4}$. This allows us to compute the $C_l$ in the flat-sky approximation \citep[e.g.][]{White:1999ApJ,Mesinger:2012MNRAS}. We calculate PS for different bandwidths $20~{\rm MHz}\lsim \Delta\nu \lsim 2.6~{\rm GHz}$, where the upper limit of the interval represents the total bandwidth of the simulation. For $\Delta \nu \lsim 1.3~{\rm GHz}$, this procedure provides multiple bandwidths that we assume to be independent to calculate the averaged $C_l$.

In upper panel of Fig. \ref{fig_ps}, we plot the power spectrum as $P_{l}=l (l+1)C_{l}/2\pi$ as a function of the angular scale ($\theta=2\pi/l$) for different bandwidths, as indicated in the legend. Independently of $\Delta\nu$, the power spectrum has a noise-like shape, i.e. $P_{l}\propto l^{2}$, for $l\lsim 10^6$. In correspondence of smaller scales, $P_l$ reaches a peak at $\theta \simeq 1\arcsec$ and then flattens, confirming the qualitative analysis discussed in Sec. \ref{sec_cmb_th_anal}.

In the central panel of Fig. \ref{fig_ps}, we compare our result ($\Delta\nu\simeq2.6$, solid black line) and PS for CMB distortions/fluctuations induced by primordial chemistry \citep[][violet dotted line]{Basu:2007NewAR}. The latter PS is obtained for \CIIion~that are assumed to be uniformly distributed ($\Delta=1$) at\footnote{As shown in \citet[][in particular, see left panel of Fig. 1]{Basu:2007NewAR}, at a different redshift, the primordial chemistry PS induced by \CIIion~ is comparable to the one reported here.} $z=4$ with $Z=10^{-1}\zsun$. With a grey shaded region we indicate the cosmic variance of the \CII-induced CMB fluctuations. As a reference, we plot the primary CMB power spectrum inferred from \emph{Planck} observations \citep[][red crosses]{Planck:2013_XVI_parameters} and calculated with \textlcsc{CMBFAST} \citep[][green dashed line]{CMBFAST:2000ApJS}.

For the primordial fluctuations, $P_{l}$ has a functional dependence on $l$ as the primary CMB power spectrum \citep[][]{Basu:2007NewAR,Schleicher:2008AA}. With respect to the primary power spectrum, these induced fluctuations produce a $\sim10^{4}$ smaller power, e.g. $P_{l}\lsim \,\mu {\rm K}^{2}$ on scales $l\lsim 10^{3}$ \citep{Basu:2004na,Basu:2007NewAR}.

We define, the PS peak as $\langle\Delta T\rangle=\sqrt{P_l}$ at $\theta\simeq 1\arcsec$, and we study its dependence on $\Delta\nu$. The result is shown in the lower panel of Fig. \ref{fig_ps}, where we plot the PS peak mean value (red circles), the r.m.s. variance (green pentagons) and the max/min values (blue triangles) as a function of $\Delta\nu$. We restate that the mean and variance are calculated for the multiple bandwidths extracted from the simulation: they must not be confused with variation of CMB fluctuations from different metal bubbles.

The peak value increases with decreasing bandwidth up to $\Delta\nu\simeq300\,{\rm MHz}$, and decreases for smaller $\Delta\nu$. This behavior can be explained as follows. The peak value increases as the bandwidth becomes comparable to metal bubble size in frequency space (see Fig. \ref{fig_esempio}). As we further decrease $\Delta\nu$, it becomes increasingly difficult to find enriched structure in the selected bandwidth, and the peak becomes shallower.

Finally, note that the peak ($\sqrt{P_{l}}$) increases faster than $\Delta\nu^{-1}$ for $2.6\gsim\Delta\nu/{\rm GHz}\gsim0.3$ and decreases for smaller $\Delta\nu$. The power spectrum of a pure noise is expected to behave as $\Delta\nu^{-1/2}$. Therefore -- in principle -- the peculiar trend with $\Delta\nu$ can discriminate a signal from CMB fluctuations from noise.

\label{lastpage}

\end{document}

%% file: include_tex/pacchetti.tex
\usepackage[english]{babel}
\usepackage{amsmath,amssymb}
\usepackage{graphicx, subfig}
\usepackage[normalem]{ulem}

\usepackage{verbatim}

\usepackage{color}
\usepackage{multirow}
\usepackage{mathtools}
\usepackage{epstopdf}

%% file: include_tex/journals.tex
\def\nat{Nature}

\def\mnras{MNRAS}
\def\apj{ApJ}
\def\apjl{ApJL}
\def\apjs{ApJS}
\def\aap{A\&A}

\def\araa{ARA\&A}
\def\aj{AJ}

\def\physrep{Phys. Rep.}
\def\pasp{Publ. Astr. Soc. Pac.}

\def\ssr{Space Science Reviews}

\def\nar{New Astronomy Reviews}

%% file: include_tex/definizioni.tex
\def\be{\begin{equation}} 
\def\ee{\end{equation}} 
\def\ba{\begin{eqnarray}} 
\def\ea{\end{eqnarray}}

\def\msun{{\Msun}}

\def\HII{\hbox{H~$\scriptstyle\rm II\ $}}

\def\CIV{\hbox{C~$\scriptstyle\rm IV\ $}}

\def\gsim{\lower.5ex\hbox{\gtsima}} 
\def\lsim{\lower.5ex\hbox{\ltsima}} \def\gtsima{$\; \buildrel > \over 
\sim \;$} \def\ltsima{$\; \buildrel < \over \sim \;$} \def\prosima{$\; 
\buildrel \propto \over \sim \;$} \def\gsim{\lower.5ex\hbox{\gtsima}} 
\def\lsim{\lower.5ex\hbox{\ltsima}} 
\def\simgt{\lower.5ex\hbox{\gtsima}} 
\def\simlt{\lower.5ex\hbox{\ltsima}} 
\def\simpr{\lower.5ex\hbox{\prosima}} \def\la{\lsim} \def\ga{\gsim} 
  
 \def\gtsima{$\; \buildrel > \over \sim \;$} 
\def\ltsima{$\; \buildrel < \over \sim \;$} 
\def\gsim{\lower.5ex\hbox{\gtsima}} 
\def\lsim{\lower.5ex\hbox{\ltsima}} 
\def\simgt{\lower.5ex\hbox{\gtsima}} 
\def\simlt{\lower.5ex\hbox{\ltsima}} 
\def\simpr{\lower.5ex\hbox{\prosima}} 
\def\la{\lsim} 
\def\ga{\gsim}

\def\msun{\,{\rm \Msun}}

\def\E3{{\cal E}_{\rm g}^{III}}

\def\r12{r_{1/2}} 
\def\x12{x_{1/2}} 
\def\v12{v_{1/2}}

%% file: include_tex/additional_def.tex
\def\diinu{\Delta I_{\nu}\slash I_{\nu}}
\def\diiDnu{\Delta I\slash I}

\def\mujb{\mu {\rm Jy}/{\rm beam}}

\def\msun{{\rm M}_{\odot}}
\def\zsun{{\rm Z}_{\odot}}

\def\CIV{\hbox{C~$\scriptstyle\rm IV $}} 

\def\SiI{\hbox{Si~$\scriptstyle\rm I $}}
\def\CII{\hbox{[C~$\scriptstyle\rm II $]}}
\def\CIIion{\hbox{C~$\scriptstyle\rm II $}}
 
\def\OI{\hbox{[O~$\scriptstyle\rm I $]}} 
\def\OIII{\hbox{[O~$\scriptstyle\rm III $]}} 
\def\NII{\hbox{[N~$\scriptstyle\rm II $]}}

\newcommand\textlcsc[1]{\textsc{\MakeLowercase{#1}}}
\newcommand{\quotes}[1]{``#1''}

\def\arcsec{^{\prime\prime}}